\documentclass[twocolumn,superscriptaddress,letter]{revtex4-2}%
\usepackage{eurosym}
\usepackage{amsbsy}
\usepackage{latexsym,epsfig,graphicx}
\usepackage{dcolumn}
\usepackage{subfigure}
\usepackage{comment}
\usepackage{multirow}
\usepackage{color}
\usepackage{bm}
\usepackage{mathrsfs}
\usepackage{amssymb}
\usepackage{amsfonts}
\usepackage{amsmath}
\usepackage{xspace}
\usepackage{soul}
\usepackage{cancel}
\usepackage{float}
\usepackage{epstopdf}
\usepackage{tabularx}
\usepackage{longtable}
\usepackage[colorlinks=true, letterpaper=true, pdfstartview=FitV, linkcolor=red, citecolor=blue, urlcolor=blue]{hyperref}
\usepackage[normalem]{ulem}
\usepackage[version=4]{mhchem}
\begin{document}
\title{Scale-tailored localization and its observation in non-Hermitian electrical circuits}
\author{Cui-Xian Guo}
\thanks{These authors contributed equally to the work}
\affiliation{Beijing National Laboratory for Condensed Matter Physics, Institute of Physics, Chinese Academy of Sciences, Beijing 100190, China}
\affiliation{Beijing Key Laboratory of Optical Detection Technology for Oil and Gas, China University of Petroleum-Beijing, Beijing 102249, China}
\affiliation{Basic Research Center for Energy Interdisciplinary, College of Science, China University of Petroleum-Beijing, Beijing 102249, China}
\author{Luhong Su}
\thanks{These authors contributed equally to the work}
\affiliation{Beijing National Laboratory for Condensed Matter Physics, Institute of Physics, Chinese Academy of Sciences, Beijing 100190, China}
\affiliation{School of Physical Sciences, University of Chinese Academy of Sciences, Beijing 100049, China}
\author{Yongliang Wang}
\affiliation{CAS Center for Excellence in Superconducting Electronics, Shanghai Institute of Microsystem and Information Technology, Chinese Academy of Sciences, Shanghai 200050, China}
\author{Li Li}
\affiliation{Beijing National Laboratory for Condensed Matter Physics, Institute of Physics, Chinese Academy of Sciences, Beijing 100190, China}
\affiliation{School of Physical Sciences, University of Chinese Academy of Sciences, Beijing 100049, China}
\author{Jinzhe Wang}
\affiliation{Beijing National Laboratory for Condensed Matter Physics, Institute of Physics, Chinese Academy of Sciences, Beijing 100190, China}
\author{Xinhui Ruan}
\affiliation{Beijing National Laboratory for Condensed Matter Physics, Institute of Physics, Chinese Academy of Sciences, Beijing 100190, China}
\affiliation{Department of Automation, Tsinghua University, Beijing 100084, China}
\author{Yanjing Du}
\affiliation{Beijing National Laboratory for Condensed Matter Physics, Institute of Physics, Chinese Academy of Sciences, Beijing 100190, China}
\author{Dongning Zheng}
\email{dzheng@iphy.ac.cn}
\affiliation{Beijing National Laboratory for Condensed Matter Physics, Institute of Physics, Chinese Academy of Sciences, Beijing 100190, China}
\affiliation{School of Physical Sciences, University of Chinese Academy of Sciences, Beijing 100049, China}
\affiliation{Hefei National Laboratory, Hefei 230088, China}
\author{Shu Chen}
\email{schen@iphy.ac.cn}
\affiliation{Beijing National Laboratory for Condensed Matter Physics, Institute of Physics, Chinese Academy of Sciences, Beijing 100190, China}
\affiliation{School of Physical Sciences, University of Chinese Academy of Sciences, Beijing 100049, China}
\author{Haiping Hu}
\email{hhu@iphy.ac.cn}
\affiliation{Beijing National Laboratory for Condensed Matter Physics, Institute of Physics, Chinese Academy of Sciences, Beijing 100190, China}
\affiliation{School of Physical Sciences, University of Chinese Academy of Sciences, Beijing 100049, China}

\begin{abstract}
Anderson localization and non-Hermitian skin effect are two paradigmatic wave localization phenomena, resulting from wave interference and the intrinsic non-Hermitian point gap, respectively. In this study, we unveil a novel localization phenomenon associated with long-range asymmetric coupling, termed scale-tailored localization, where the number of induced localized modes and their localization lengths scale exclusively with the coupling range. We show that the long-range coupling fundamentally reshapes the energy spectra and eigenstates by creating multiple connected paths on the lattice. Furthermore, we present experimental observations of scale-tailored localization in non-Hermitian electrical circuits utilizing adjustable voltage followers and switches. The circuit admittance spectra possess separate point-shaped and loop-shaped components in the complex energy plane, corresponding respectively to skin modes and scale-tailored localized states. Our findings not only expand and deepen the understanding of peculiar effects induced by non-Hermiticity but also offer a feasible experimental platform for exploring and controlling wave localizations.
\end{abstract}
\maketitle

The recent surge of research in non-Hermitian physics \cite{colladd3,nhreview,nhreview2,coll6,nhreview3,nhreview4} has uncovered a wide array of phenomena that transcend the realm of traditional Hermitian systems. Non-Hermitian systems exhibit a remarkable sensitivity to their boundary conditions, exemplified by the non-Hermitian skin effect (NHSE) \cite{nhse1,nhse3,nhse4,nhse5,nhse6,nhse7,nhse8,nhse9,pointtopo3,ddnhse,ddnhse_hu,ddfangchen}. Featured by the gathering of a significant number of eigenstates at system boundaries, the NHSE breaks the extended Bloch-wave behaviors and challenges the conventional notion of bulk-edge correspondence by displaying distinct spectral shapes under periodic and open boundary conditions \cite{nhsee1,nhsee2,nhsee3,nhsee4,graph1,graph2}. In addition to the skin effect, the interplay between non-Hermiticity and spatial inhomogeneity, such as domain walls, disorders, or impurities/defects, offers intriguing insights into wave behaviors and introduces additional richness to localization phenomena in generic non-Hermitian systems. They include impurity-induced topological bound states \cite{defect1,defect2}, non-Hermitian quasi-crystals \cite{nhquasicrystal1,nhquasicrystal2}, and the counterintuitive accumulation of eigenstates known as scale-free localization \cite{sfl_li,GuoCX,sfl_guo,sfl_wang,sflerg,sdse_guo}, where eigenstates concentrate near defects while the localization length scales with the entire system size. Electric circuits \cite{LeeCH,circuitB4,circuitB5,circuitB9,nhsee4} offer a powerful platform for simulating various lattice models, where the lattice Hamiltonian is represented by an adjacency matrix that adheres to Kirchhoff's law in current networks. Non-reciprocity can be modeled using active devices\cite{Hofmann}, facilitating the realization of the NHSE in topolectric circuits\cite{nhsee3}. The key asymmetric element is implemented using a negative impedance converter through current inversion, known as the INIC device.

\begin{table}[tb]
\caption{Comparison of four distinct wave-localization phenomena based on localization length, position, physical origin, and fate in the thermodynamic limit. They include (1) Anderson localization; (2) NHSE: non-Hermitian skin effect; (3) SFL: scale-free localization; and (4) STL: scale-tailored localization. For Anderson localization or NHSE, the localization length or skin depth is finite. While for SFL and STL, the localization length scales with the total system size and coupling range, respectively.}\label{table1}
\begin{center}
\renewcommand{\arraystretch}{1.5}
\begin{tabular}{|c|c|c|c|c|}
\hline
&Anderson & NHSE & SFL & STL\\
\hline
Loc. length & $\mathcal{O}(1)$ & $\mathcal{O}(1)$ & $\mathcal{O}(N)$ & $\mathcal{O}(l)$\\
\hline
\multirow{2}{*}{Position} & \multirow{2}{*}{bulk} & \multirow{2}{*}{boundary} & \multirow{2}{*}{impurity} & boundary\\
& & & & /impurity \\
\hline
\multirow{2}{*}{Origin} & wave & point & local&long-range \\
    & interference & gap & impurity  & coupling \\
\hline
Theo. limit & \checkmark & \checkmark & $\times$ & \checkmark \\
\hline
\end{tabular}
\end{center}
\end{table}

The skin localization and the remarkable spectral sensitivity have been harnessed for innovative functionalities such as optical funneling \cite{funneling}, high-precision sensor devices \cite{sensor1,circuitB9,sensor2}, and optomechanically induced transparency \cite{optintran}. Yet, the collective localization of all eigenstates inherent in the NHSE poses the challenging task of engineering a uniform distribution of spatial non-Hermiticity across the entire system. Furthermore, it restricts the tunable freedom of wave dynamics such as localization length, position and proportion of eigenstates. The pivotal question is: Can we precisely tailor wave localization in a controllable manner within generic lattice systems?

In this work, we uncover a novel type of localization of eigenstates, termed scale-tailored localization (STL) [See Table \ref{table1}].
Unlike Anderson localization resulting from wave interference or NHSE arising from intrinsic point gaps \cite{nhse5,nhse6,ddnhse_hu}, STL emerges due to the presence of long-range asymmetric coupling, giving rise to localized modes with distinctive characteristics: both their number and localization length scale exclusively with the coupling range. Consequently, the energy spectra are partitioned into two distinct sectors: point- (or arc-) shaped spectra corresponding to modes barely affected by the long-range coupling, and loop-shaped spectra associated with the scale-tailored localized states. This is in stark contrast to the NHSE, where skin localization requires a uniform distribution of non-Hermiticity (e.g., gain/loss or nonreciprocity) across the entire lattice, and the skin depths of the eigenstates are fixed and governed by the non-Bloch band theory \cite{nhse1,nhse3}. From a practical standpoint, STL customizes wave localization without the complexity of a full-scale implementation of non-Hermiticity. The scale-tailored localized modes, unlike the scale-free localized modes induced by local non-Hermiticity \cite{sfl_guo, sfl_wang, sflerg}, exhibit resilience in the thermodynamic limit.

In our setup, we implement a unidirectional electrical circuit with a rolled boundary condition controlled by electric switches, where unidirectional coupling is achieved through a simplified version of active devices, the voltage followers (VFs) \cite{sdse_guo}. The scale-tailored localized modes can be transformed into skin modes or vice versa with the changing of the asymmetric coupling range, accompanied by the self-adaptation of the localization length of all scale-tailored localized states. We then identify the STL by measuring the admittance spectra in the non-Hermitian electrical circuits. Our results indicate that long-range asymmetric coupling (non-Hermiticity) can serve as a powerful tool to manipulate wave localization and offer a feasible platform for exploring the intriguing properties of scale-tailored localized states.\\

\begin{figure}[t]
\includegraphics[width=0.46\textwidth]{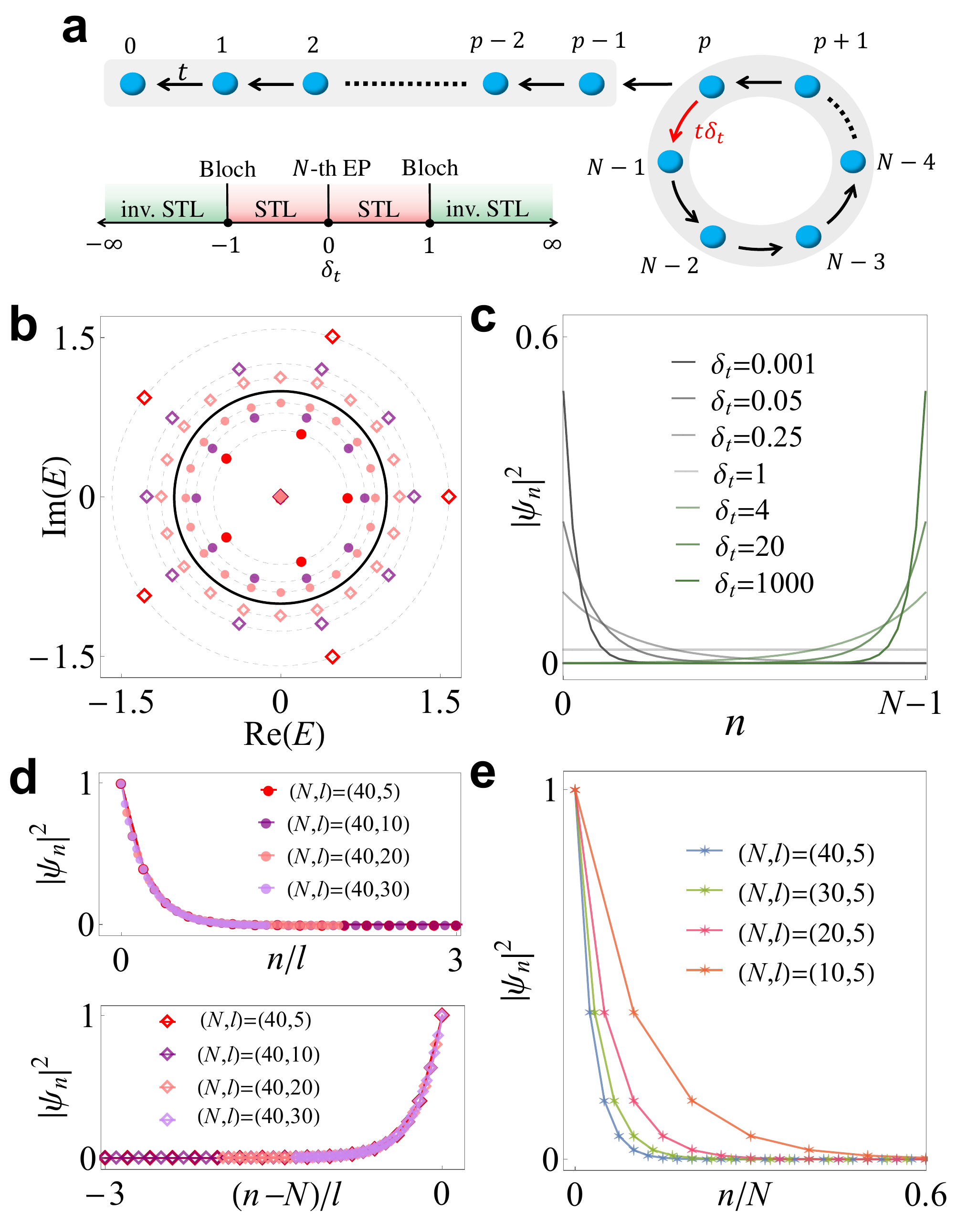}
\caption{Scale-tailored localization (STL) in the unidirectional hopping model. (a) Upper panel: Sketch of the lattice model with rolled boundary condition. Bottom panel: Different regimes (marked in different colors) of eigenstates' localization with respect to the coupling strength $\delta_t$. STL (inv. STL): the induced second-type eigenstates are localized on the left (right) boundary; Bloch: all second-type eigenstates are extended. (b) Eigenenergies for different combinations of $(N,l)$ in the complex plane with $\delta_t=0.1$ (solid circles) or $\delta_t=10$ (empty diamonds). The case of $(N,l)=(40,5)$, $(40,10)$, $(40,20)$ are marked in red, purple, and pink, respectively. For reference, the black unit circle represents the spectra under periodic boundary condition. (c) Spatial distributions of the second-type eigenstates for several typical values of $\delta_t$ with fixed $(N,l)=(40,20)$. (d) The perfect overlapping of rescaled spatial distributions by the coupling range $l$ for different $(N,l)$. (e) Rescaled spatial distributions by the total system size $N$ with $\delta_t=0.1$ and $l=5$.}
\label{fig1}
\end{figure}

\noindent{\large{\textbf{Results}}}\\
\textbf{Scale-tailored localization}

\noindent We start by introducing a minimal model that exhibits STL. It has unidirectional hoppings on a one-dimensional (1D) lattice with rolled boundary condition, as depicted in Fig. \ref{fig1}(a). The Hamiltonian for this model is expressed as:
\begin{equation}\label{HE}
\hat{H}=\sum\limits_{n=0}^{N-2}t \hat{c}_n^{\dag }\hat{c}_{n+1}+t \delta_t \hat{c}_{N-1}^{\dag }\hat{c}_p.
\end{equation}
Here, $N$ represents the length of the lattice. $\hat{c}_n^{\dag }$ and $\hat{c}_n$ are the creation and annihilation operators on the $n$-th site. $t$ is the strength of the unidirectional hopping between neighboring sites and set to be the energy unit $t=1$ in the following. Our model features an additional long-range asymmetric hopping connecting the end site and the inner $p$-th site with coupling strength $\delta_t$. The coupling range is $l=N-p$. The special case of $p=0$ and $\delta_t=1$ or $\delta_t=0$ corresponds to the periodic or open boundary condition.

In the presence of the additional coupling, the eigenvalues and eigenstates of the system are
\begin{equation}\label{Solution}
\left\{
\begin{array}{ll}
E=z, \\
|\Psi\rangle=(\psi_0,\psi_1,\cdots ,\psi_{N-1})^{T}=(1,z,\cdots ,z^{N-1})^{T},
\end{array}
\right.
\end{equation}
with $z$ given by the roots of the following equation:
\begin{equation}\label{Solution-z}
z^{p}\left(\delta_t -z^{N-p}\right)=0.
\end{equation}
The energy spectra are highly sensitive to the boundary conditions as evident from the solutions. For periodic boundary condition ($\delta_t=1$ and $p=0$), the solutions are $z^{(m)}=e^{i\frac{2\pi}{N}m}$ ($m=0,1,...,N-1$), and all eigenstates are extended Bloch states. The eigenenergies are uniformly distributed on the unit circle. In contrast, for open boundary condition ($\delta_t=0$), there exists a unique $N$-fold degenerate solution $E=0$. All eigenstates coalesce into the state $|\Psi\rangle=(1,0,...,0)^T$ residing at the first site of the lattice. This degeneracy arises from $N\times N$ Jordan-block form of the Hamiltonian, which leads to an $N$-th order exceptional point (EP). In our subsequent studies, we focus on the more general cases with $\delta_t\neq0$ and $p\neq0$. The solutions can be classified into two distinct types. The first type corresponds to a $p$-fold degenerate solution $z=0$, representing a $p$-th order EP with eigenstates localized exclusively at the first site. The second type consists of $l$ non-degenerate solutions given by $z^{(m)}=\sqrt[l]{\delta_t}e^{i\theta_m}$, where $m=1,2,\cdots,l$, $\theta_m=\frac{2\pi}{l}m$, and $l$ is the number of rolled sites. Interestingly, the first-type solutions can be regarded as remnants of the $N$-th order EP, unaffected by the additional long-range coupling. In contrast, the second-type solutions result from such long-range coupling, with eigenenergies and wavefunctions given by:
\begin{equation}\label{Psi2}
\left\{
\begin{array}{ll}
E_m = \sqrt[l]{\delta_t}e^{i\theta_m}, \\
|\Psi^{(m)}\rangle = \left(1, \sqrt[l]{\delta_t}e^{i\theta_m}, \cdots , \left(\sqrt[l]{\delta_t}e^{i\theta_m}\right)^{N-1}\right)^{T}.
\end{array}
\right.
\end{equation}

These $l$ eigenenergies evenly distributed on a circle of radius $\sqrt[l]{|\delta_t|}$, dependent solely on the settings of the long-range coupling. The entire energy spectra are composed of both the isolated EP at the center and loop-shaped parts circling around, separated in the complex plane, as shown in Fig. \ref{fig1}(b).

We proceed to examine the localization properties of the second-type eigenstates in Eq. (\ref{Psi2}). They have the same spatial profile but different site-dependent phase factors. The localization length $\xi$ can be extracted from the spatial profile via $|\psi^{(m)}_n|\sim e^{-\frac{|n-n_0|}{\xi}}$, with $n_0$ the localization center. When the additional coupling is weaker than the unidirectional hopping $|\delta_t|<1$, all $l$ states accumulate at $n_0=0$ with localization length
\begin{equation}
\xi=-\frac{l}{\log|\delta_t|}.
\end{equation}
While if the coupling is stronger than the unidirectional hopping $|\delta_t|>1$, they accumulate at the last site $n_0=N-1$ with localization length $\xi=\frac{l}{\log|\delta_t|}$. Figure \ref{fig1}(c) illustrates the eigenstates' profiles for several typical values of $\delta_t$ with fixed $(N,l)=(40,20)$. The more $|\delta_t|$ deviates from 1, the stronger the localization becomes. A duality exists between $|\delta_t|$ and $|\frac{1}{\delta_t}|$ with equal localization length but opposite localization directions. At $\delta_t=1$, they become extended across the whole lattice. The different localization regimes of the second-type eigenstates are summarized in Fig. \ref{fig1}(a).

The analysis above clearly indicates that even an infinitesimal long-range coupling can trigger eigenstates' localization, underscoring its non-perturbative nature. Intriguingly, the localization length of the second-type eigenstates is directly proportional to the coupling range $l$, and is irrelevant to the total system size $N$, which differs from the scale-free localization \cite{sfl_li,sfl_guo,sfl_wang,sflerg}. Upon rescaling these eigenstates by the coupling range, their spatial profiles become perfectly identical, as depicted in Fig. \ref{fig1}(d). We refer to this type of accumulation as STL. The case $|\delta_t|>1$ is termed inverse STL due to the opposite localization direction. In comparison, Fig. \ref{fig1}(e) shows the eigenstates' rescaling with respect to the total system $N$. In the thermodynamic limit, the scale-free localized modes become extensive, while the scale-tailored localized modes maintain a fixed and finite localization length.\\

\begin{figure}[tbp]
\includegraphics[width=0.48\textwidth]{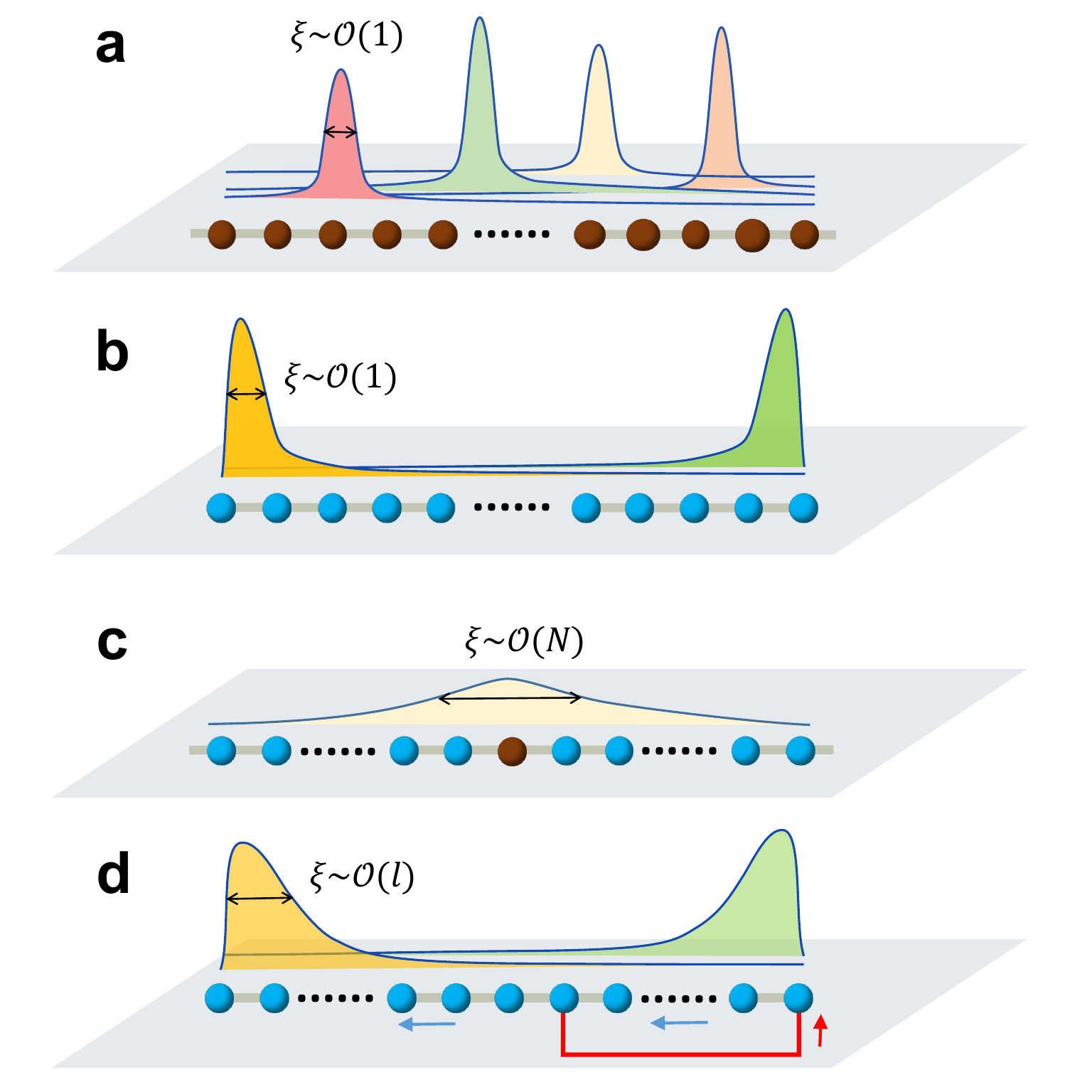}
\caption{Schematics of several typical localization phenomena: (a) Anderson localization in disordered lattice resulting from wave interference; (b) Non-Hermitian skin modes localized at the system boundary due to nontrivial point gap; (c) Scale-free localization induced by a local non-Hermitian impurity; (d) Scale-tailored localization (STL) arising from long-range asymmetric coupling.}
\label{fig2}
\end{figure}

\noindent\textbf{The mechanism and generality of STL}\\
The STL exhibits unique characteristics that set it apart from other localization phenomena, as outlined in Table \ref{table1}. Figure \ref{fig2} sketches the physical mechanism of several typical localizations. In Anderson localized systems, the eigenstates have finite localization lengths due to wave interference in disordered media, impeding wave propagation. The non-Hermitian skin effect, scale-free localization, and STL are specific to non-Hermitian systems. For the skin effect, the skin modes are confined to the system boundary, with finite localization lengths governed by the generalized Brillouin zone. The skin effect accompanies spectral collapses from Bloch bands of periodic boundary conditions and requires the intrinsic point-gap topology or spectral winding. In contrast, the scale-free localization is the eigenstates' accumulation near an impurity, with their localization length proportional to the system length, $\xi\sim\mathcal{O}(N)$. In the simplest scenario, the presence of a local non-Hermitian impurity gives rise to an $\mathcal{O}(1/N)$ correction to the eigenspectra and eigenstates. In STL, the long-range coupling introduces a new length scale $l$. Heuristically, the long-range coupling can be treated as a non-local non-Hermitian impurity that generates closed paths within the one-dimensional lattice. The wave propagation at the junction (e.g., the $p$-th lattice site of model (\ref{HE})), satisfies a self-consistency condition $f(z^l,\delta_t)=0$ (e.g., $z^l=\delta_t$ in Eq. (\ref{Solution-z})). While the specific form of $f(z^l,\delta_t)$ depends on model details, the self-consistency condition yields $l$ states characterized by a localization length of $\xi \sim 1/\log|z| \sim \mathcal{O}(l)$. The highly size-dependent spectral properties of the critical NHSE \cite{llh_nc,qinfang_prb} can be visualized from the extreme case of STL when $l=N$ with $N$ the system length.

While we have presented the simplest unidirectional coupling model for illustrative purposes, it is worth noting that the phenomenon of STL induced by long-range asymmetric coupling is expected to be quite general and applicable to other models as well, such as the Hatano-Nelson model and Hermitian lattice chain (details in Supplementary Sections I,II and Figs. S1-S4). This holds true regardless of whether the original system possesses skin modes or not. Moreover, scale-tailored localized states can emerge both at the system boundary and in the vicinity of long-range impurities (details in Supplementary Section III and Fig. S5). The STL persists in the presence of multiple long-range asymmetric couplings (details in Supplementary Section IV and Figs. S6,S7). These couplings induce various scale-tailored eigenstates of different length scales. We have further verified the occurrence of STL in 2D  (details in Supplementary Section V and Figs. S8,S9) and in interacting systems \cite{nhse_interacting1,nhse_interacting2,Rshen_interacting3} (details in Supplementary Section VI and Fig. S10).
The presence of STL, which involves the reshaping of a fraction of the eigenspectra and eigenstates (scaling with the coupling range $l$), implies the potential for effectively manipulating wave localization by suitably tailoring the long-range couplings in the system.

Now, let us consider the most general case with lattice Hamiltonian
\begin{equation}\label{GHm}
\hat{H}=\sum\limits_{n=0}^{N-j-1}\left[ \sum\limits_{j=1}^{M_L}t_{jL}\hat{c}_n^{\dag}\hat{c}_{n+j}+\sum\limits_{j=1}^{M_R}t_{jR}\hat{c}_{n+j}^{\dag}\hat{c}_n \right]+\delta_t \hat{c}_{N-1}^{\dag }\hat{c}_p,
\end{equation}
where $t_{jL}$ ($t_{jR}$) represents the hopping towards the left (right) side, with the largest range being $M_L$ ($M_R$). $\delta_t$ denotes the asymmetric coupling with range $l=N-p$. We can analytically solve the eigenspectra and eigenstates of Hamiltonian (\ref{GHm}) and establish a general criterion for the occurrence of STL (details in the Methods, Supplementary Section II and Figs. S3,S4). We take the Bloch spectra under periodic boundary conditions $E=\sum\limits_{j=1}^{M_R}\frac{t_{jR}}{z^{j}}+\sum\limits_{j=1}^{M_L}t_{jL}z^{j}$. For a given $E$ inside the Bloch spectra, there exist $M=M_R+M_L$ solutions $z_i~(i=1,\cdots,M)$, which can be ordered by their moduli $|z_1|\le|z_2|\le\cdots\le|z_M|$. Note that there must exist solutions with unit modulus because $E$ is chosen from the Bloch spectra. If the $(M_R+1)$-th root has unit modulus:
\begin{eqnarray}\label{criterion}
|z_{M_R}|<|z_{M_R+1}|=1,
\end{eqnarray}
then there are $l$ scale-tailored states when adding an asymmetric coupling.\\

\begin{figure*}[t]
\includegraphics[width=0.95\textwidth]{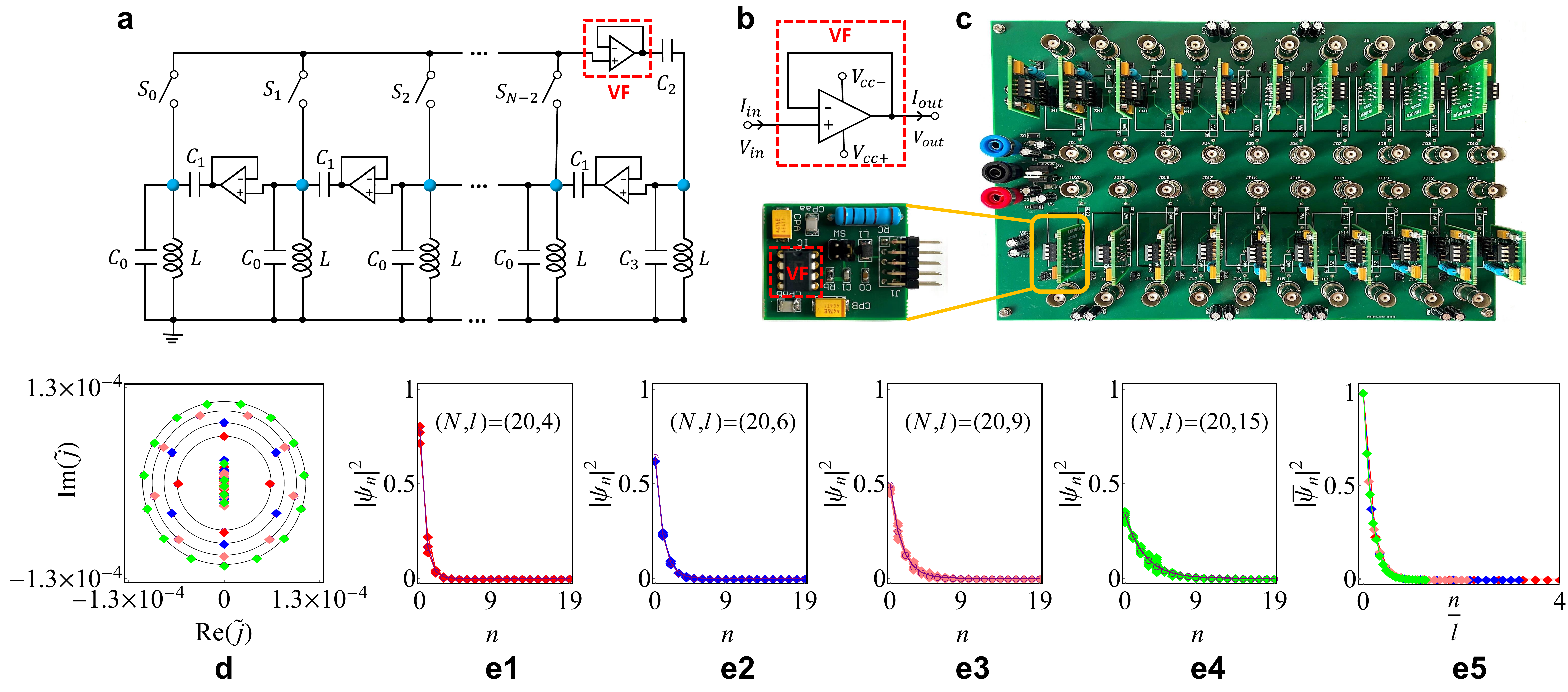}
\caption{Implementation of unidirectional electrical circuit and observation of STL. (a) Experimental design of the unidirectional circuit array. Each unit cell consists of an LC resonator with a capacitor $C_0=10$pF and an inductor $L=220\mathrm{\mu H}$, a VF, and a connecting capacitor $C_1=220$pF. (b) VF as the active device with mismatched input and output currents. (c) The fabricated circuit board with 20 unit cells. The long-range coupling is controlled by electrical switches. (d) Eigenvalues of the admittance matrix with different $(N,l)=(20,4)$ (red), $(20,6)$ (blue), $(20,9)$ (pink), and $(20,15)$ (green). (e1-e4) Spatial profiles of all scale-tailored localized eigenstates (corresponding to the eigenvalues distributed on the circle) of the admittance matrix with different $(N,l)$. The experimental results (solid diamonds) align with the theoretical results (hollow circles). (e5) Rescaled spatial distributions divided by the coupling range $l$ of the scale-tailored localized eigenstates. Other parameters are: $\omega=2\pi\times100$kHz, $C_2=10$pF, $C_3=220$pF.}
\label{fig3}
\end{figure*}

\noindent\textbf{Unidirectional electrical circuit}\\
We implement the unidirectional-hopping model using electrical circuit that combines passive and active devices, as illustrated in Figs. \ref{fig3}(a-c). Each unit cell in the circuit comprises an LC resonator with a capacitor $C_0=10$pF (except for the last one with $C_3$) and an inductor $L=220\mathrm{\mu H}$, a VF, and a connecting capacitor $C_1=220$pF which couples two neighboring nodes. To achieve the rolled boundary condition, or the long-range coupling, we activate the $p$-th switch while leaving the other switches off. The coupling strength in the long bond is controlled by capacitor $C_2$. The key element responsible for unidirectionality is the VF with mismatched input and output currents. This active device utilizes an operational amplifier (OpAmp) to replicate the input voltage at the output, as depicted in Fig. \ref{fig3}(b). The voltage or current at the input and output ends satisfy the relation:
\begin{equation} \begin{split} V_{\text{out}} = V_{\text{in}},~~~I_{\text{in}} = 0. \end{split} \end{equation}

A printed circuit board comprising 20 units is fabricated, as displayed in Fig. \ref{fig3}(c). Based on Kirchhoff's law, for a given alternating current (AC) input current with frequency $\omega$, the circuit lattice is described by
\begin{equation}
\mathbf{I}(\omega)=J(\omega)\mathbf{V}(\omega),
\end{equation}
where $J$ represents the admittance matrix (or the circuit Laplacian). The current and voltage vectors are defined as $\mathbf{I}=(I_0,I_{1},\cdots,I_{N-1})$ and $\mathbf{V}=(V_0,V_{1},\cdots,V_{N-1})$, respectively, with $I_n$ and $V_n$ denoting the input current and voltage at node $n$. The admittance matrix $J$ and its eigenvalues play a role similar to the Hamiltonian (up to some trivial constant term) and its associated eigenenergies. In our circuit, the capacitor at the last unit is set to $C_3=220$pF, and the coupling capacitor is $C_2=10$pF, corresponding to $\delta_t=0.0454$ in the unidirectional model (\ref{HE}). The AC input current has a frequency of $\omega=2\pi f=2\pi\times 100$kHz, and the coupling range $l$ is adjustable using switches.\\

\noindent\textbf{Experimental demonstration of STL}\\
By measuring the voltage responses at all nodes in the network when subjected to a local current input, the admittance eigenvalues and eigenstates can be accessed. To demonstrate the scaling rule with the coupling range, we examine four representative cases: $l=N-p=4, 6, 9, 15$, while maintaining a fixed lattice size of $N=20$ by activating the corresponding switches. In Fig. \ref{fig3}(d), we present the measured admittance spectra in the complex energy plane. For each $l$, the spectra consist of two distinct types: $l$ states evenly distributed on a circle with a radius of $(\delta_t)^{1/l}$, and the remaining $N-l$ states enclosed within this circle. These $N-l$ states primarily reside at the first site, arising from the $p$-fold degenerate exceptional point, which is highly sensitive to perturbations. Imperfections or non-uniformities in the capacitors/inductors can cause the exceptional point to split and spread along the imaginary axis (details in Supplementary Section VII, Fig. S11 and Table S1). In contrast, the states distributed on the circle display resilience.

In Figs. \ref{fig3}(e1)-(e4), we present the spatial profiles of eigenstates (with their corresponding eigenvalues distributed on the circle) of the admittance matrix for different combinations of $(N,l)$. Notably, for each $l$, the eigenstates display nearly identical profiles, with small deviations due to the unavoidable circuit noises or device errors. These eigenstates decay from the left boundary exponentially with a finite spanning. As the coupling range $l$ increases, the eigenstates gradually become more delocalized. Furthermore, by rescaling their distributions with the prefactor of $l$, we observe perfect overlapping of their profiles for all combinations of $(N,l)$, as depicted in Fig. \ref{fig3}(e5). It indicates that the localization length scales as $\mathcal{O}(l)$, thus confirming the nature of the scale-tailored localized states.

\begin{figure}[tbp]
\includegraphics[width=0.45\textwidth]{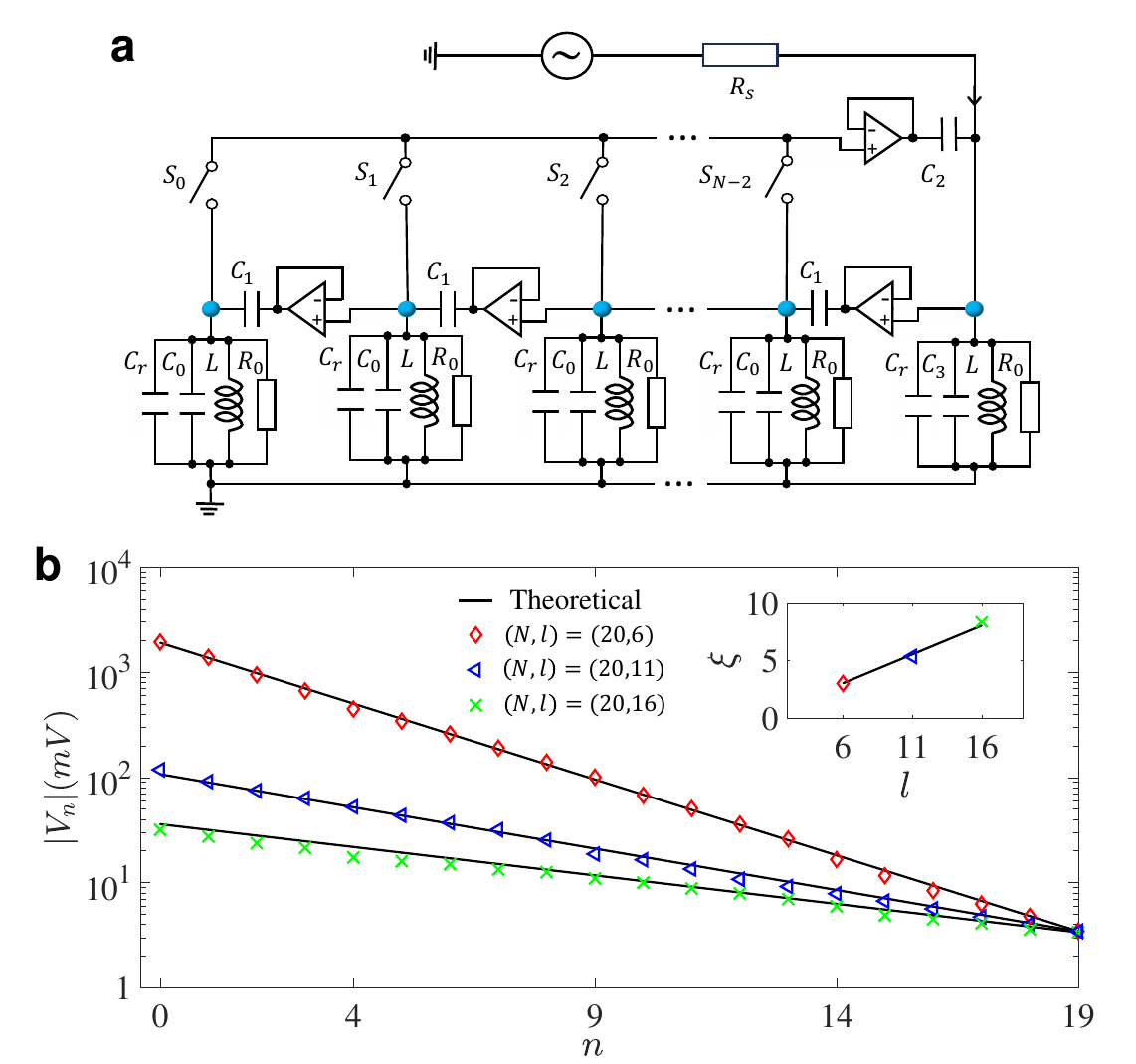}
\caption{Direct measurement of the scale-tailored localized states. (a) An input AC current is fed at the far right end of the circuit using an AC voltage source connected through a shunt resistance $R_s=10k\Omega$. (b) Voltage response at all nodes relative to the current feed. The inset displays the localization length $\xi$ extracted from the voltage responses. Experimental data are marked by colored symbols for different configurations: $(N,l)=(20,6)$ (red diamonds) with $(R_0,f)=(82k\Omega,174$kHz), $(20,11)$ (blue triangles) with $(R_0,f)=(22k\Omega,175$kHz), $(20,16)$ (green crosses) with $(R_0,f)=(16k\Omega,176.5$kHz), which align with theoretical expectation (black lines). Other parameters are: $C_0=10$pF, $C_1=220$pF $C_2=30$pF, $C_3=200$pF, $C_r=1.5$nF, $L=470\mu$ H.}
\label{fig4}
\end{figure}

The reconstruction of the scale-tailored states from the admittance matrix involves $N^2$ voltage response measurements, which is cumbersome and indirect. Instead, these eigenstates can be obtained through the measurement of the non-local voltage response, as illustrated in Fig. \ref{fig4}(a). An AC current feed is injected at the far right end of the circuit using an AC voltage source connected through a shunt resistance $R_s=10k\Omega$. Each additional resistor $R_0$ and capacitor $C_r$ per unit cell is used only for circuit stability and facilitating frequency adjustment. When the driving frequency approaches the system's eigenfrequency corresponding to a scale-tailored eigenstate, the measured voltage response directly yields the profile of the scale-tailored eigenstate [See Methods]. Intriguingly, despite the current being fed at the far right end, the measured voltage response peaks strongest at the far left end, as shown in Fig. \ref{fig4}(b). This counterintuitive enhancement underscores the exotic localization behavior of the scale-tailored eigenstate. In fact, the localization length $\xi$ can be extracted from the voltage response:
\begin{equation}
\xi=\frac{1}{\log\left[\frac{1}{N-1}\sum_{n=0}^{N-2}\frac{|V_n|}{|V_{n+1}|}\right]},
\end{equation}
where $V_n$ represents the voltage response at $n$-th node. As shown in the inset of Fig. \ref{fig4}(b), the linear scaling of $\xi$ with the coupling range $l$ is further confirmed.\\

\noindent{\large{\textbf{Discussion}}}

\noindent We have established STL as a novel localization mechanism stemming from long-range asymmetric couplings, going beyond the well-known paradigms of Anderson localization due to wave interference and skin localization arising from intrinsic non-Hermitian point gaps. Leveraging the high feasibility of electric-circuit arrays and the adjustability of nonreciprocity through VFs, we have further observed the scale-tailored localized states in electrical circuits, accompanied by the separation of energy spectra in the complex plane. Our framework highlights the non-perturbative nature of non-Hermitian couplings, resulting in dramatic changes in energy spectra and eigenstates. With wave localization fully tailored by the long-range coupling, our study opens new avenues for the versatile manipulations of peculiar wave phenomena in various open systems and other experimental platforms, including photonic \cite{nhsee2, funneling}, ultracold atoms \cite{nhsecold}, and metamaterials \cite{nhsee1}.\\

\noindent{\large{\textbf{Methods}}}\\
\textbf{Analysis of STL}\\
The unidirectional-hopping model in Eq. (\ref{HE}) is a special case of the more generic Hatano-Nelson model \cite{HNPRL} with nonreciprocal couplings:
\begin{equation}\label{HE2}
\hat{H}=\sum\limits_{n=0}^{N-2}(t_L\hat{c}_n^{\dag}\hat{c}_{n+1}+t_R \hat{c}_{n+1}^{\dag }\hat{c}_{n})+t_L \delta_t \hat{c}_{N-1}^{\dag }\hat{c}_p.
\end{equation}
Here, $t_R$ and $t_L$ represent the hopping to the right and left site, respectively. The eigenvalues and eigenstates for the above model can be obtained as:
\begin{equation}\label{Solution2}
\left\{
\begin{array}{ll}
E=t_Lz_i+\frac{t_R}{z_i}, \\
|\Psi\rangle=\sum\limits_{i=1}^{2}|\Psi_i\rangle=\sum\limits_{i=1}^{2}c_i(1,z_i,z_i^{2},\cdots ,z_i^{N-1})^{T},
\end{array}
\right.
\end{equation}
where $z_1$ and $z_2$ satisfy $z_1z_2=t_R/t_L=\eta$, and $z_1$ is given by the roots of the following equation:
\begin{equation}\label{EqZ2}
z_1^{N+1}-\left(\frac{\eta}{z_1}\right)^{N+1}+\delta_t\left[\left(\frac{\eta}{z_1}\right)^{p+1}-z_1^{p+1}\right]=0.
\end{equation}
Without the additional long-range coupling, i.e., $\delta_t\mapsto 0$, Eq. (\ref{EqZ2}) reduces to $z_1^{2(N+1)}=\eta^{(N+1)}$, from which we obtain $N$ solutions $z_{1/2}^{(m)}=\sqrt{\eta}e^{\pm i\theta_m}$ with $\theta_m=[m\pi/(N+1)]~(m=1,\cdots,N)$. These roots form the generalized Brillouin zone \cite{nhse1}. The eigenstates are skin modes localized at the left or right boundary when $|t_R|<|t_L|$ or $|t_R|>|t_L|$, respectively.

When $\delta_t$ deviates from $0$ and exceeds a critical value, part of the skin modes are reshaped by the long-range coupling. For simplicity, we focus on the case of $N,p\gg l$ $(l=N-p)$, that is, the coupling range $l$ is the smallest length scale of the system. When $\delta_t>r^l$, the solutions of Eq. (\ref{EqZ2}) can be categorized into two types. The first type consists of $p$ solutions satisfying $|z_1|=|z_2|=\sqrt{\eta}$. Specifically, they are $z_{1/2}=\sqrt{\eta}e^{\pm i\theta}$,  where $\theta$ is determined by real solutions of $r^{l}\sin[(N+1)\theta]=\delta_t\sin[(p+1)\theta]$. These solutions represent the skin modes that are nearly unaffected by the long-range coupling. The second type has $l$ solutions satisfying $|z_2|<\sqrt{\eta}<|z_1|$. In this case, Eq. (\ref{EqZ2}) reduces to $z_1^{N-p}=\delta_t$, which leads to $l$ solutions $z_1^{(m)}=\sqrt[l]{\delta_t}e^{i\theta_m}$ with $\theta_m=\frac{2m\pi}{l}$ $(m=1,2,\cdots,l)$. The localization length of the second-type eigenstates are determined by the settings of the long-range coupling. They correspond to STL or inverse STL for $|\delta_t|<1$ or $|\delta_t|>1$ with localization length $\xi=\mp\frac{l}{\log|\delta_t|}\propto l$.\\

\begin{figure}[!t]
\includegraphics[width=0.45\textwidth]{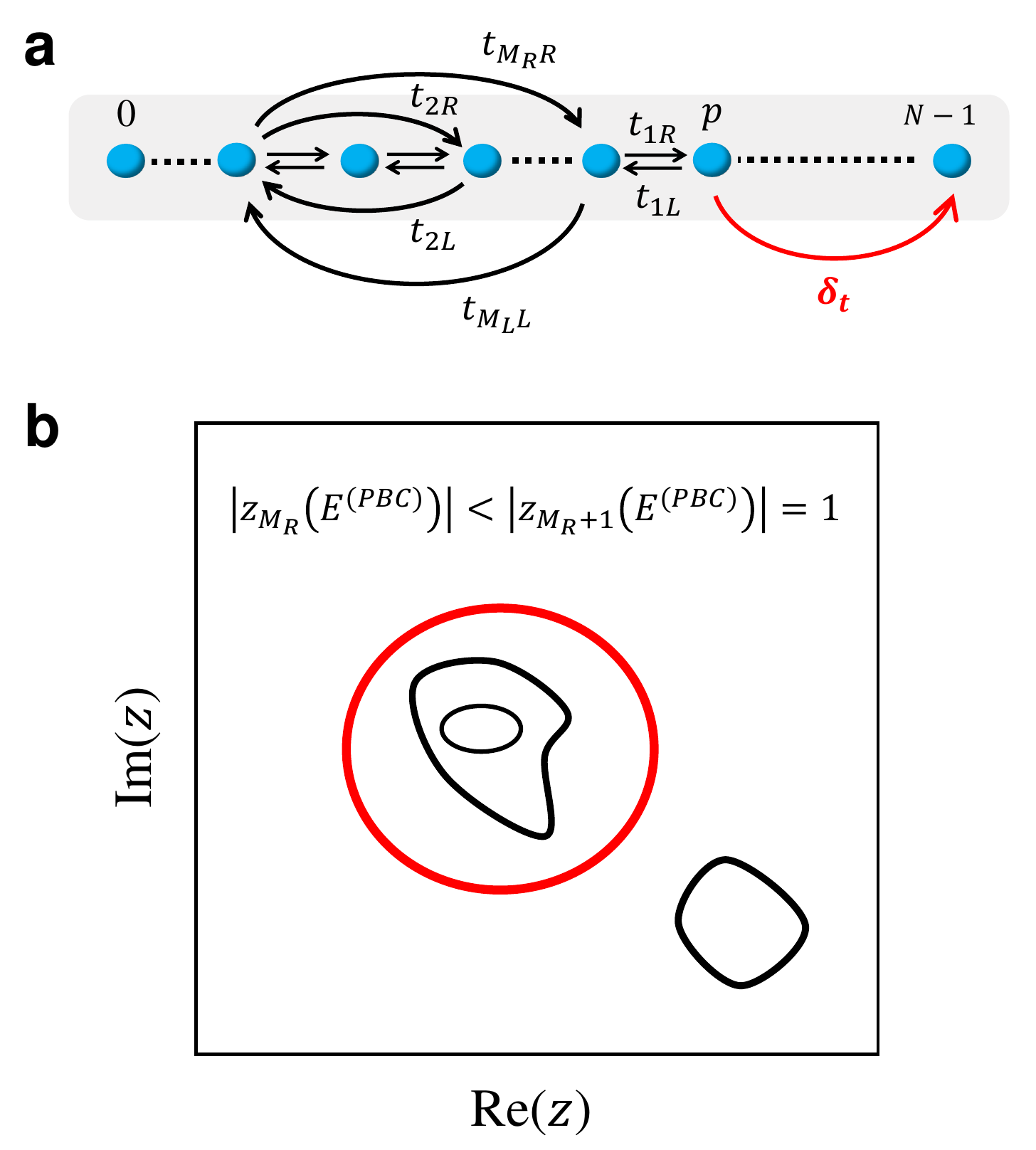}
{\caption{Sketch of generic 1D lattice models and the criterion of STL. (a) Sketch of the 1D generic non-Hermitian lattice model. The largest hopping ranges to the left and right are $M_L$ and $M_R$. The additional asymmetric coupling is marked in red. (b) Sketch of the criterion of STL. The $(M_R+1)$-th roots of the Bloch spectra reside on the unit circle (red circle).}
\label{fig5}}
\end{figure}

\noindent\textbf{The criterion of STL}\\
Here, we investigate the generic non-Hermitian model described by Hamiltonian (\ref{GHm}) with an additional non-local coupling, as sketched in Fig. \ref{fig5}(a). The solution of the eigenvalue equation $\hat{H}|\Psi\rangle = E|\Psi\rangle$ are
\begin{equation}
\left\{
\begin{array}{ll}
E = \sum\limits_{j=1}^{M_R} \frac{t_{jR}}{z^j} + \sum\limits_{j=1}^{M_L} t_{jL} z^j,\\
|\Psi\rangle=\sum\limits_{i=1}^{M}c_i(1,z_i,z_i^{2},\cdots ,z_i^{N-1})^{T}.
\end{array}
\right.
\end{equation}
Here $M = M_R + M_L$, and $c_1, \cdots, c_M$ are superposition coefficients determined by the boundary constraints $\det[H_B] = 0$. For a given $E$, there exist $M$ solutions $z_i~(i=1,\cdots,M)$, which can be ordered as $|z_1| \le |z_2| \le \cdots \le |z_M|$.

We focus on the relevant case with $N \gg l \gg 1$ and discuss the existence condition of STL (details in Supplementary Section II). The dominant terms in the determinant are $\det[H_B] = A_1 + A_2 + B_1$, with
\begin{equation}
\begin{split}
A_1 =& \left( z_{M_R+1} z_{M_R+2} \cdots z_M \right)^N G_a \\
& \times \sum_{i_1 \neq \cdots \neq i_{M_R}=1}^{M_R} (-1)^{\tau(i_1 \cdots i_{M_R})} \left[ f_1(z_{i_1}) \cdots f_{M_R}(z_{i_{M_R}}) \right], \\
\end{split}
\end{equation}
\begin{equation}
\begin{split}
A_2=&\left(z_{M_R}z_{M_R+2}\cdots z_{M}\right)^{N}G'_a\\
&\times\sum_{i_1 \neq \cdots \neq i_{M_R}=1}^{M_R-1,M_R+1}(-1)^{\tau(i_1\cdots i_{M_R})}\left[f_1(z_{i_1})\cdots f_{M_R}(z_{i_{M_R}})\right],\\
\end{split}
\end{equation}
\begin{equation}
\begin{split}
B_1 =& -\delta_t z_{M_R+1}^p \left( z_{M_R+2} \cdots z_M \right)^N G_b \\
& \times \sum_{i_1 \neq \cdots \neq i_{M_R}=1}^{M_R} (-1)^{\tau(i_1 \cdots i_{M_R})} \left[ f_1(z_{i_1}) \cdots f_{M_R}(z_{i_{M_R}}) \right].
\end{split}
\end{equation}
Here, $G_a$, $G'_a$ and $G_b$ are finite polynomials depending on the specific model. If $|B_1| \gg |A_2|$, which requires
\begin{equation}\label{Gcondition}
\delta_t |z_{M_R+1}|^p \gg |z_{M_R}|^N,
\end{equation}
then the boundary constraints yield $A_1 + B_1 = 0$. Consequently,
\begin{equation}\label{Gzsolu}
z_{M_R+1} = \sqrt[l]{\delta_t \eta} e^{i \theta_m},
\end{equation}
where
\begin{equation}
\eta = \frac{G_b}{G_a},
\end{equation}
and $\theta_m = \frac{2m\pi}{l}$ with $m = 1, \cdots, l$. Since $G_a$ and $G_b$ are finite polynomials, $|z_{M_R+1}|^l = |\delta_t \eta| \sim \mathcal{O}(1)$, indicating there are $l$ scale-tailored localized states.

Substituting Eq. (\ref{Gzsolu}) into Eq. (\ref{Gcondition}), the condition (\ref{Gcondition}) simplifies to
\begin{equation}\label{Gdtt}
\delta_t > |z_{M_R}|^{l}.
\end{equation}
This governs the existence of $l$ scale-tailored localized states with $|z_{M_R+1}| = \sqrt[l]{\delta_t \eta}$ under the condition $N \gg l \gg 1$. Notably, for sufficiently large $l$, $|z_{M_R+1}| \rightarrow 1$, and the eigenenergies of these scale-tailored states approach the Bloch spectra (energy spectra under periodic boundary conditions):
\begin{equation}\label{1}
E = E^{(PBC)} = \sum_{j=1}^{M_R} \frac{t_{jR}}{(e^{ik})^j} + \sum_{j=1}^{M_L} t_{jL} (e^{ik})^j,
\end{equation}
where $k \in [0, 2\pi]$. Moreover, the condition Eq. (\ref{Gdtt}) holds regardless of the magnitude of $\delta_t$ only if $|z_{M_R}|<1$.
Thus, the criterion can be formulated in terms of the Bloch spectra: for any $E \in \sigma_{PBC}$, there exist $M_L + M_R$ solutions $z_i$, sorted by their moduli $\left|z_1\right| \le \left|z_2\right| \le \ldots \le \left|z_{M_L+M_R}\right|$. Note that for the Bloch spectra, there must exist a $z$-solution of unit modulus. If these solutions $z_i$ further satisfy $\left|z_{M_R}\right| < \left|z_{M_R+1}\right| = 1$, then $l$ scale-tailored localized states appear upon introducing the additional long-range coupling of arbitrary strength, as sketched in Fig. \ref{fig5}(b).\\

\noindent\textbf{Circuit Laplacian and impedance matrix}\\
The circuit Laplacian relates the input current and voltage at all the nodes via Kirchhoff's law, $\mathbf{I}(\omega)=J(\omega)\mathbf{V}(\omega)$. For the circuit array shown in Fig. \ref{fig3}(a) and Fig. \ref{fig4}(a), the circuit Laplacian is given by:
\begin{equation}\label{Jw}
\begin{split}
  J(\omega)&=-i\omega\left(
      \begin{array}{cccccccc}
        \mu & C_1 & \cdots & 0 & 0 & \cdots & 0 & 0 \\
        0 & \mu & \cdots & 0 & 0 & \cdots & 0 & 0 \\
        \cdots & \cdots & \cdots & \cdots & \cdots & \cdots & \cdots & \cdots \\
        0 & 0 & \cdots & \mu & C_1 & \cdots & 0 & 0 \\
        0 & 0 & \cdots & 0 & \mu & \cdots & 0 & 0 \\
        \cdots & \cdots & \cdots & \cdots & \cdots & \cdots & \cdots & \cdots \\
        0 & 0 & \cdots & 0 & 0 & \cdots & \mu & C_1 \\
        0 & 0 & \cdots & C_2 & 0& \cdots & 0 & \mu \\
      \end{array}
    \right)\\
    &=\widetilde{J}(\omega)-i\omega\mu I_{N\times N}.
\end{split}
\end{equation}
Here $\mu=\frac{1}{\omega^2 L}-(C_1+C_0)$ in Fig. \ref{fig3}, and $\mu=\frac{1}{\omega^2 L}-(C_1+C_0+C_r)-\frac{1}{i\omega R_0}$ in Fig. \ref{fig4}. The capacitors $C_0$ and $C_3$ are set to satisfy $C_1+C_0=C_2+C_3$ in our circuit. $I_{N\times N}$ is the $N\times N$ identity matrix. $J_{N-1,p}=-i\omega C_2$ represents the long-range coupling. Compared to the theoretical model in Eq. (\ref{HE}), we have the coupling strength $t$ and $t\delta_t$ set by the capacitors $C_1$ and $C_2$ in the circuit array. In the experiments illustrated in Fig. \ref{fig3}, each unit cell is composed of an LC resonator, a capacitor, and a VF (OpAmp OP07) with a gain bandwidth product of $600$ kHz. In Fig. \ref{fig4}, each unit cell incorporates an additional resistor $(R_0)$ and a capacitor $(C_r)$, with the VF (OpAmp OP27G) having an $8$ MHz bandwidth product. We plug the circuit unit into the circuit motherboard, allowing for easy adjustment of both boundary coupling and lattice size. To mitigate crosstalk between adjacent inductors, we maintain a distance of approximately $4$ cm between two lattice sites.

To access the admittance eigenvalues and eigenstates, we perform voltage-response measurements with respect to a local current input for all nodes in the network. These responses are encoded in the impedance matrix $G$:
\begin{equation}
\mathbf{V}(\omega)=G(\omega)\mathbf{I}(\omega).
\end{equation}
Specifically, with an input AC current $I_n$ at the $n$-th node and the measured voltage response $V_{m}^{n}$ at the $m$-th node, the impedance matrix element $G_{mn}$ is given by:
\begin{equation}
G_{mn}=\frac{V_{m}^{n}}{I_{n}}.
\end{equation}
The admittance matrix $J(\omega)$ and the impedance matrix are related through $J(\omega)=G^{-1}(\omega)$. In the experimental setup shown in Fig. \ref{fig3}, the AC current is provided by an AC voltage source (NF Wave Factory1974) through a resistance of $R_s=2k\Omega$, and the voltage response is measured using a lock-in amplifier (Zurich Instruments UHF).

Besides the reconstruction of the admittance matrix, a direct access of the scale-tarilored eigenstates is possible through the non-local voltage measurements as in Fig. \ref{fig4}(a). The impedance matrix $G\left(\omega\right)$ (the inverse of the admittance matrix $J\left(\omega\right))$ encodes information about the eigenmodes. It can be expressed as
\begin{equation}
G\left(\omega\right)=
\begin{pmatrix}
G_0 & G_1 \\
0 & G_2 \\
\end{pmatrix},
\end{equation}
where $G_0$ is a $p\times p$ upper triangular matrix with elements being $[G_0]_{i,j}=\frac{-1}{J_{i,i+1}}\left(\frac{-J_{i,i+1}}{J_{i,i}}\right)^{j-i+1}$. $G_1$ is a $p\times l$ matrix defined by $[G_1]_{i,j}=\left(\frac{-J_{i,i+1}}{J_{i,i}}\right)^{p-i}[G_2]_{0,j}$. $G_2$ is an $l\times l$ matrix:
\begin{equation}
G_2=\sum_{n=p}^{N-1}{\frac{1}{j_n}\frac{{\psi_{nR}^\prime\psi}_{nL}^{\prime\dag}}{\psi_{nL}^{\prime\dag}\psi_{nR}^\prime}}
\end{equation}
with $\psi_{nR,i}^\prime=\psi_{nR,p+i}$ and $\psi_{nL,i}^\prime=\psi_{nL,p+i}$. Here, $\psi_{nR}$ or $\psi_{nL}~(n=p,\cdots,N-1)$ is the right or left eigenvector with eigenenergy $j_n$ of the admittance matrix $J\left(\omega\right)$, representing the scale-tailored eigenstates. We denote the eigenfrequency of the electric circuits as $\omega_c^{\left(m\right)}$, determined by $\det\left[J\left(\omega_c^{\left(m\right)}\right)\right]=0$. When the driving frequency approaches an eigenfrequency $\omega_c^{\left(m\right)}$, the eigenvalue associated with a scale-tailored eigenstate satisfies $j_m\left(\omega\rightarrow\omega_c^{\left(m\right)}\right)\rightarrow0$ and $J_{i,i}=-j_m$. We thus have
\begin{equation}
\left[G_1\left(\omega\rightarrow\omega_c^{\left(m\right)}\right)\right]_{i,j}=\frac{1}{j_m}\frac{\psi_{mL,p+j}^\ast}{\psi_{mL}^{\prime\dag}\psi_{mR}^\prime}\psi_{mR,i}
\end{equation}
with $i=0,\cdots,p-1$, and $j=0\cdots,l-1$. The matrix $G_2$ reduces to
\begin{equation}
\left[G_2\left(\omega\rightarrow\omega_c^{\left(m\right)}\right)\right]_{i,j}=\frac{1}{j_m}\frac{\psi_{mL,p+j}^\ast}{\psi_{mL}^{\prime\dag}\psi_{mR}^\prime}\psi_{mR,p+i}
\end{equation}
with $i=0,\cdots,l-1$, and $j=0\cdots,l-1$. This indicates that when the driving frequency approaches the eigenfrequency $\omega_c^{\left(m\right)}$, the $x$-th $\left(x=p,\cdots,N-1\right)$ column of the impedance matrix directly yields the scale-tailored eigenstate $\psi_{mR}$, i.e.,
\begin{equation}
\left[G\left(\omega\rightarrow\omega_c^{\left(m\right)}\right)\right]_x=\frac{1}{j_m}\frac{\psi_{mL,x}^\ast}{\psi_{mL}^{\prime\dag}\psi_{mR}^\prime}\psi_{mR}\sim\psi_{mR}.
\end{equation}
Therefore, the scale-tailored eigenstates can be accessed by measuring the voltage response related to the input AC current at the far-right end of the circuit under $\omega \rightarrow \omega_c^{(m)}$.\\

\noindent {\large{\textbf{Data availability}}}\\
The data used in this study are available in the GitHub repository \href{https://github.com/G-CX1/STL-Code}{https://github.com/G-CX1/STL-Code}.\\

\noindent {\large{\textbf{Code availability}}}\\
The code used in this study is available in the GitHub repository \href{https://github.com/G-CX1/STL-Code}{https://github.com/G-CX1/STL-Code}.\\

\noindent {\large{

\vspace{0.5cm}

\begin{acknowledgments}
\noindent {\large{\textbf{Acknowledgments}}}\\
This work is supported by National Key Research and Development Program of China (Grant No. 2023YFA1406704 and Grant No. 2022YFA1405800), the NSFC under Grants No. 12174436, No. T2121001, and No. 92265207, Innovation Program for Quantum Science and Technology (Grant No. 2021ZD0301800), and the Strategic Priority Research Program of Chinese Academy of Sciences under Grants No. XDB33000000 and No. XDB28000000. C.-X. G. is also supported by the China Postdoctoral Science Foundation (No. 2024M753608) and Science Foundation of China University of Petroleum, Beijing (No. 2462024SZBH003).\\
\end{acknowledgments}

\noindent {\large{\textbf{Author contributions}}}\\
H. H., S. C., and D. Z. conceived the work; C.-X. G. did the major part of the theoretical derivation and numerical calculation; L. S, C.-X. G., Y. Wang, L. Li, J. Wang, X. Ruan, and Y. Du conducted the experiments and analyzed the data; All authors discussed the results and participated in the writing of the manuscript.\\

\noindent {\large{\textbf{Competing interests}}}\\
The authors declare no competing interests.\\

\onecolumngrid
\newpage
\renewcommand{\theequation}{S\arabic{equation}}
\renewcommand{\thefigure}{S\arabic{figure}}
\renewcommand{\thetable}{S\arabic{table}}
\setcounter{equation}{0}
\setcounter{figure}{0}
\setcounter{table}{0}

\begin{center}
    {\bf \large Supplementary Material for ``Scale-tailored localization and its observation in non-Hermitian electrical circuits" }
\end{center}

\noindent This supplementary material provides details on:\\
\noindent (I) The scale-tailored localization (STL) in the Hatano-Nelson model and Hermitian lattice chain;\\
\noindent (II) The criterion of STL and more examples;\\
\noindent (III) Scale-tailored localized states in the vicinity of impurities;\\
\noindent (IV) Scale-tailored localized states induced by multiple long-range asymmetric couplings;\\
\noindent (V) The STL in the 2D unidirectional hopping model;\\
\noindent (VI) The STL in the Bose-Hubbard model;\\
\noindent (VII) Analysis of experimental imperfection or non-uniformity in capacitors and inductors.\\

\section*{(I) The STL in the Hatano-Nelson model and Hermitian lattice chain}
The emergence of STL induced by long-range asymmetric coupling is not limited to the unidirectional hopping model. Here, we investigate the more generic Hatano-Nelson model with Hamiltonian
\begin{equation}\label{SHHN}
\hat{H}=\sum\limits_{n=0}^{N-2}(t_L\hat{c}_n^{\dag}\hat{c}_{n+1}+t_R \hat{c}_{n+1}^{\dag }\hat{c}_{n})+t_L \delta_t \hat{c}_{N-1}^{\dag }\hat{c}_p,
\end{equation}
where $t_R$ and $t_L$ represent the hopping amplitudes towards the right and left site, respectively. $\delta_t$ denotes the long-range asymmetric coupling from $p$-th site to the last site, with a length of $l=N-p$. The unidirectional hopping model studied in the main text is a special case of the Hatano-Nelson model with $t_R=0$. The eigenvalue equation for model (\ref{SHHN}) is
\begin{equation}
\hat{H}|\Psi\rangle=E|\Psi\rangle,
\end{equation}
with $|\Psi\rangle=\sum_{n}\psi _{n}c_{n}^{\dag}|0\rangle$. We take an ansatz wave function $\Psi(z)=(1,z,\cdots ,z^{N-1})^{T}$. For bulk lattice sites,
\begin{equation}\label{SH3BE}
t_R\psi _{n}-E\psi _{n+1}+t_L\psi _{n+2}=0,
\end{equation}
with $n=0,1,\cdots,N-3$. For boundary lattice sites,
\begin{equation}\label{SH3BdE}
\begin{split}
-E\psi _{0}+t_L \psi _{1}&=0,\\
t_L\delta_t \psi _{p}+t_R \psi _{N-2}-E\psi _{N-1}&=0.\\
\end{split}
\end{equation}
By substituting the ansatz into Eqs. (\ref{SH3BE}), we obtain the eigenvalues
\begin{equation}\label{SH3E}
E=\frac{t_R}{z}+t_Lz.
\end{equation}
Given a value of $E$, there exist two solutions $z_i~(z_1,z_2)$ that satisfy the following constraint:
\begin{equation}
z_1z_2=\frac{t_R}{t_L}.
\end{equation}
Therefore, the general solution of the eigenfunction should be the superposition
\begin{equation}\label{SH3wvg}
(\psi_0,\psi_1,\cdots,\psi_{N-1})^T=c_1\Psi(z_1)+c_2\Psi(z_2).
\end{equation}
Let us further insert Eq. (\ref{SH3wvg}) and Eq. (\ref{SH3E}) into boundary equations Eq. (\ref{SH3BdE}), the superposition coefficients $c_1$ and $c_2$ fulfill
\begin{equation}\label{XSbb1}
H_{B}\left(
       \begin{array}{c}
         c_{1} \\
         c_{2} \\
       \end{array}
     \right)
=0,
\end{equation}
with
\begin{equation}
H_{B}=\left(
\begin{array}{cc}
z_1^{-1} & z_2^{-1} \\
\delta_tz_1^p-z_1^N &  \delta_tz_2^p-z_2^N
\end{array}%
\right).
\end{equation}
The nontrivial solutions of $(c_1,c_2)$ are determined by $\det[H_B]=0$, yielding the constraint for $z_i$:
\begin{equation}\label{SH3EqZ12}
z_1^{N+1}-z_2^{N+1}+\delta_t\left[z_2^{p+1}-z_1^{p+1}\right]=0.
\end{equation}
For convenience, we set $z_{1/2}=re^{\pm i\theta}$ with $r=\sqrt{t_R/t_L}$, $f_1=r^{l}\sin[(N+1)\theta]$, $f_2=\delta_t\sin[(p+1)\theta]$. The solutions of $\theta\in \mathbb{C}$ are determined by
\begin{equation}\label{Sf12}
f_1=f_2.
\end{equation}
\begin{figure}[!h]
\includegraphics[width=0.9\textwidth]{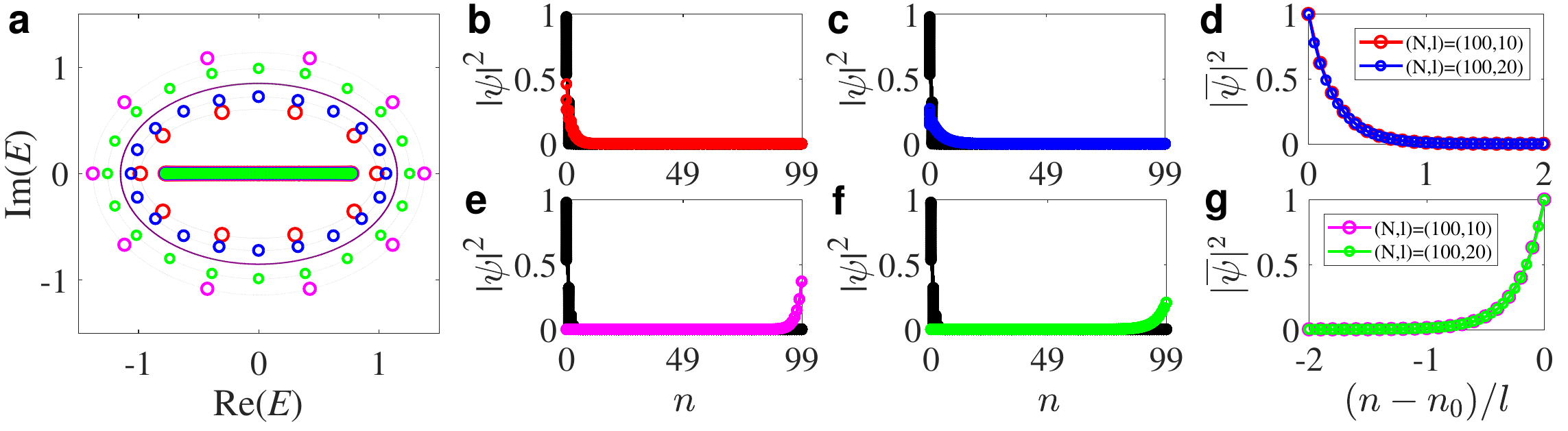}
\caption{Eigenenergies and eigenfunctions for the Hatano-Nelson model with $t_R/t_L=0.15$. (a) Eigenenergies for different $(N,l,\delta_t)$ in the complex plane. The case of $(N,l,\delta_t)=(100,10,0.1)$, $(N,l,\delta_t)=(100,20,0.1)$, $(N,l,\delta_t)=(100,10,10)$ and $(N,l,\delta_t)=(100,20,10)$ are marked in red, blue, magenta and green, respectively. For reference, the purple circle represents the spectra under periodic boundary condition. (b,c,e,f) Eigenfunctions for different $(N,l,\delta_t)$. The skin modes are marked in black, while the (inverse) scale-tailored localized states are marked in red, blue, magenta and green, respectively. (d,g) Rescaled spatial distributions by the coupling range $l$ of the (inverse) scale-tailored localized eigenstates for $\delta_t=0.1$ and $\delta_t=10$, respectively.}
\label{figSM1}
\end{figure}

We first consider the case of $t_R\neq t_L$. The real roots of $\theta$ correspond to skin modes due to $|z_i|=r\neq 1$. Without the additional long-range coupling, i.e. $\delta_t=0$, all $N$ eigenfunctions represent skin modes with $z_{1,2}=re^{\pm\frac{ im\pi}{(N+1)}}~(m=1,\cdots,N)$. With the increase of the asymmetric coupling strength $\delta_t$, the number of skin modes $N_{skin}$ gradually decreases until it reaches the minimum value of $N_{skin}=p$. To be explicit,
\begin{equation}\label{SNskin}
\begin{split}
N_{skin}=
\left\{
  \begin{array}{ll}
    N,~~~~~~~~~~\rm{for}~\hbox{$0\leq\delta_t\leq \delta_{t_a}$};\\
    p\sim N,~~~~\rm{for}~\hbox{$\delta_{t_a}<\delta_t<\delta_{t_b}$};\\
    p,~~~~~~~~~~~\rm{for}~\hbox{$\delta_t\geq\delta_{t_b}$}.\\
  \end{array}
\right.
\end{split}
\end{equation}
Here $\delta_{t_a}=r^{l}$ and $\delta_{t_b}=\frac{N+1}{p+1}r^{l}$ obtained from $Max(|f_1|)=Max(|f_2|)$ and $\frac{\partial f_2}{\partial\theta}|_{\theta=0}=\frac{\partial f_1}{\partial\theta}|_{\theta=0}$, respectively. In the thermodynamic limit but with finite-range asymmetric coupling, $N,p\gg l$, Eq. (\ref{SNskin}) reduces to
\begin{equation}
\begin{split}
N_{skin}=
\left\{
  \begin{array}{ll}
    N,~~~~~~~~~~\rm{for}~\hbox{$0\leq\delta_t\leq r^{l}$};\\
    p,~~~~~~~~~~~\rm{for}~\hbox{$\delta_t> r^{l}$}.\\
  \end{array}
\right.
\end{split}
\end{equation}
Obviously, for the case of $r=0$ as discussed in the main text, the number of skin modes reduces to $p$, while the remaining $l$ eigenstates fall into scale-tailored localized states even for an infinitesimally long-range coupling $\delta_t$. When $r\neq 0$ and $\delta_t>r^{l}$, the eigenvalues and eigenfunctions for the $p$ skin modes (with real $\theta$) are given by
\begin{equation}
E=2\sqrt{t_Rt_L}\cos\theta,
\end{equation}
\begin{equation}
\Psi=\left(1-e^{2i\theta},re^{i\theta}\left(1-e^{-i4\theta}\right),\cdots,r^{N-1}e^{i(N-1)\theta}\left(1-e^{-i2N\theta}\right)\right)^T.
\end{equation}
These skin modes are localized at the boundary with real eigenvalues. For the remaining $l$ states, $r$ lies in the middle of $|z_1|$, $|z_2|$ (suppose $|z_1|>|z_2|$) and $\theta$ is complex. When $N,p\gg l$, Eq. (\ref{SH3EqZ12}) reduces to
\begin{equation}
z_1^{N+1}-\delta_t z_1^{p+1}=0,
\end{equation}
by keeping the two dominant terms. Thus, we have $z_1^{(m)}=\sqrt[l]{\delta_t}e^{i\theta_m}$ and $z_2^{(m)}=r^2e^{-i\theta_m}/\sqrt[l]{\delta_t}$ with $\theta_m=\frac{2m\pi}{l}~(m=1,2,\cdots,l)$. The eigenvalues form an ellipse in the complex-energy plane:
\begin{equation}
E_m=\cos\theta_m\left(t_L\sqrt[l]{\delta_t}+\frac{t_R}{\sqrt[l]{\delta_t}} \right)+i\sin\theta_m\left(t_L\sqrt[l]{\delta_t}-\frac{t_R}{\sqrt[l]{\delta_t}} \right).
\end{equation}
The localization length $\xi$ for these $l$ eigenstates are
\begin{equation}\label{SH1xi}
\begin{split}
\xi&=
\left\{
  \begin{array}{ll}
    -\frac{l}{\log|\delta_t|},~~~\hbox{$|\delta_t|<1$,} \\
    \frac{l}{\log|\delta_t|},~~~\hbox{$|\delta_t|>1$.}
  \end{array}
\right.
\end{split}
\end{equation}
The localization center is at $n_0=0$ (for $|\delta_t|<1$, left boundary) or $n_0=N-1$ (for $|\delta_t|>1$, right boundary). The localization length is proportional to the coupling range $l$, indicating that they are scale-tailored localized states or inverse scale-tailored localized states, respectively. As illustrated in Fig. \ref{figSM1}(a), the eigenvalues associated with the skin modes and the (inverse) scale-tailored localized states are separated by arc-shaped and loop-shaped patterns in the complex energy plane. In Figs. \ref{figSM1}(b,c,e,f), we plot the corresponding wave functions for different combinations of $(N,l,\delta_t)$. The scale-tailored localized (or inverse scale-tailored localized) states accumulate on the left (or right) boundary, while skin modes are localized at the left boundary. After a rescaling of the average spatial distribution of scale-tailored localized (or inverse scale-tailored localized) states by $l$, these distributions for different $(N,l)$ overlap, as shown in Fig. \ref{figSM1}(d,g).
\begin{figure}[htb]
\includegraphics[width=0.9\textwidth]{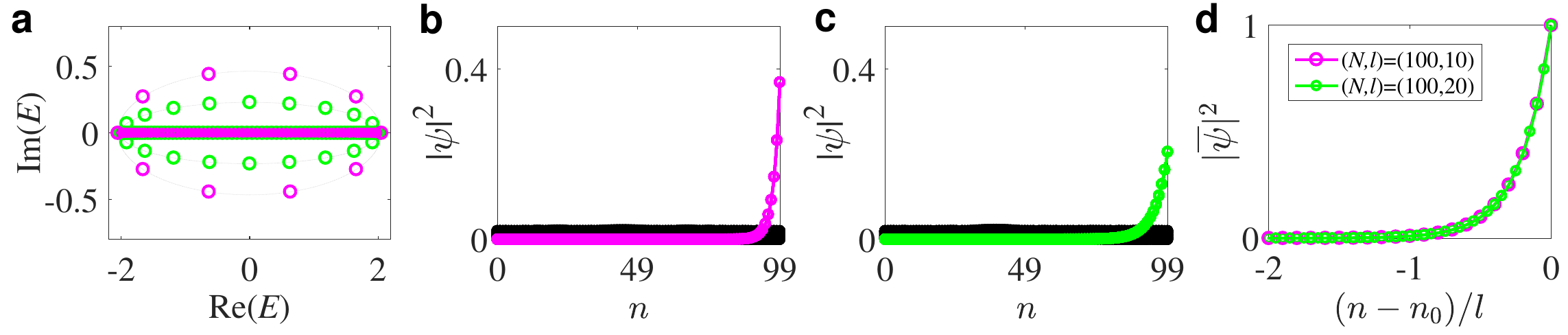}
\caption{Eigenenergies and eigenfunctions for the Hatano-Nelson model with $t_R/t_L=1$ and $\delta_t=10$. (a) Eigenenergies for different $(N,l)$ in the complex plane. The case of $(N,l)=(100,10)$ and $(N,l)=(100,20)$ are marked in magenta and green, respectively. (b,c) Eigenfunctions for different $(N,l)$. The extended states are marked in black, while the inverse scale-tailored localized states are marked in magenta and green, respectively. (d)  Rescaled average spatial distributions of the inverse scale-tailored localized eigenstates by the coupling range $l$ for different $(N,l)$.}
\label{figSM2}
\end{figure}

In the case of reciprocal hoppings for bulk lattice sites, $r=1$, the real solutions for $\theta$ in Eq. (\ref{Sf12}) correspond to extended states rather than the skin modes, and the number of these extended states is given by:
\begin{equation}
\begin{split}
N_{extended}=
\left\{
  \begin{array}{ll}
    N,~~~~~~~~~~\rm{for}~\hbox{$0\leq\delta_t\leq 1$};\\
    p,~~~~~~~~~~~\rm{for}~\hbox{$\delta_t> 1$}.\\
  \end{array}
\right.
\end{split}
\end{equation}
for $N,p\gg l$. Thus, as long as the condition $\delta_t> 1$ is satisfied, the solutions are divided into two types, one consisting of $p$ extended states and the other consisting of $l$ inverse scale-tailored localized states, as shown in Fig. \ref{figSM2}.

\section*{(II) The criterion of STL and more examples}
\subsection*{(a) The detailed derivation}
In this section, we provide a detailed derivation of the criterion for STL in generic 1D non-Hermitian models. Subsequently, we validate this criterion with several concrete examples. We consider the Hamiltonian
\begin{equation}\label{GH}
\hat{H}=\sum\limits_{n=0}^{N-j-1}\left[ \sum\limits_{j=1}^{M_L}t_{jL}\hat{c}_n^{\dag}\hat{c}_{n+j}+\sum\limits_{j=1}^{M_R}t_{jR}\hat{c}_{n+j}^{\dag}\hat{c}_n \right]+\delta_t \hat{c}_{N-1}^{\dag }\hat{c}_p,
\end{equation}
where $t_{jL}$ ($t_{jR}$) represents the hopping amplitudes towards the left (right) side. The largest hopping ranges to the left and right are  $M_L$ and $M_R$. $\delta_t$ denotes the long-range asymmetric coupling from $p$-th site to the last site with range $l=N-p$. We set $\delta_t> 0$ for convience. The eigenvalue equation for model (\ref{GH}) is
\begin{equation}
\hat{H}|\Psi\rangle=E|\Psi\rangle,
\end{equation}
with $|\Psi\rangle=\sum_{n}\psi _{n}c_{n}^{\dag}|0\rangle$. We take an appropriate ansatz wave function
\begin{equation}\label{GAns}
\Psi(z)=(1,z,\cdots ,z^{N-1})^{T}.
\end{equation}
For bulk lattice sites,
\begin{equation}\label{GBulkE}
\sum\limits_{j=1}^{M_R}t_{jR}\psi_{n-j}-E\psi _{n}+\sum\limits_{j=1}^{M_L}t_{jL}\psi_{n+j}=0.
\end{equation}
with $n=M_R,M_R+1,\cdots,N-1-M_L$. For boundary lattice sites,
\begin{equation}\label{GBdyE}
\begin{split}
-E\psi _{0}+\sum\limits_{j=1}^{M_L}t_{jL}\psi_{j}&=0,\\
\sum\limits_{j=1}^{j_a}t_{jR}\psi_{j_a-j}-E\psi _{j_a}+\sum\limits_{j=1}^{M_L}t_{jL}\psi_{j_a+j}&=0,\\
\sum\limits_{j=1}^{M_R}t_{jR}\psi_{N-1-j_b-j}-E\psi _{N-1-j_b}+\sum\limits_{j=1}^{j_b}t_{jL}\psi_{N-1-j_b+j}&=0,\\
\delta_t \psi _{p}+\sum\limits_{j=1}^{M_R}t_{jR}\psi_{N-1-j}-E\psi _{N-1}&=0.\\
\end{split}
\end{equation}
where $j_a=1,\cdots,M_R-1$ and $j_b=1,\cdots,M_L-1$.
By substituting the ansatz wave function Eq. (\ref{GAns}) into Eqs. (\ref{GBulkE}), we obtain the eigenvalues
\begin{equation}\label{GEg}
E=\sum\limits_{j=1}^{M_R}\frac{t_{jR}}{z^{j}}+\sum\limits_{j=1}^{M_L}t_{jL}z^{j}.
\end{equation}
For a given $E$, there exist $M$ ($M=M_R+M_L$) solutions $z_i$, which can be ordered by their moduli $|z_1|\le|z_2|\le\cdots\le|z_M|$. The solutions of the eigenfunction should be the superposition
\begin{equation}\label{GAnsg}
(\psi_0,\psi_1,\cdots,\psi_{N-1})^T=\sum_{i=1}^{M}c_i\Psi(z_i)=\sum_{i=1}^{M}c_i(1,z_i,z_i^2,\cdots,z_i^{N-1})^T.
\end{equation}
Let us further insert Eq. (\ref{GAnsg}) and Eq. (\ref{GEg}) into the boundary equations (\ref{GBdyE}), the superposition coefficients $c_i$ fulfill
\begin{equation}\label{XSbb1}
H_{B}\left(
       \begin{array}{c}
         c_{1} \\
         c_{2} \\
         \cdots \\
         c_{M} \\
       \end{array}
     \right)
=0,
\end{equation}
where
\begin{equation}
H_{B}=\left(
\begin{array}{ccc}
f_1(z_1) &  \cdots  &  f_1(z_{M}) \\
\vdots   &  \vdots  &  \vdots   \\
f_{M_R}(z_1) &  \cdots  &  f_{M_R}(z_{M}) \\
g_1(z_1)z_1^{N} &  \cdots  &  g_1(z_M)z_M^{N}  \\
\vdots   &  \vdots  &  \vdots   \\
g_{M_L-1}(z_1)z_1^{N} &  \cdots  &  g_{M_L-1}(z_M)z_M^{N}  \\
\left[g_{M_L}(z_1)-\frac{\delta_t}{z_1^l}\right]z_1^{N} &  \cdots  &  \left[g_{M_L}(z_M)-\frac{\delta_t}{z_M^l}\right]z_M^{N}  \\
\end{array}
\right).
\end{equation}
with
\begin{equation}
\begin{split}
&f_s(z_i)=\sum\limits_{j=s}^{M_R}\frac{t_{jR}}{z_i^{j-s+1}},~~(s=1,\cdots,M_R);\\
&g_s(z_i)=\sum\limits_{j=M_L-s+1}^{M_L}t_{jL}z_i^{j-(M_L-s+1)},~~(s=1,\cdots,M_L-1);\\
&g_{M_L}(z_i)=\sum\limits_{j=1}^{M_L}t_{jL}z_i^{j-1}.
\end{split}
\end{equation}
The non-trivial solutions $c_i~(i=1,\cdots,M)$ are determined by $\det[H_B]=0$, which expands to
\begin{equation}
\det[H_B]=\det[H_B^a]+\det[H_B^b]=0,
\end{equation}
where
\begin{equation}
H_{B}^a=\left(
\begin{array}{ccc}
f_1(z_1) &  \cdots  &  f_1(z_{M}) \\
\vdots   &  \vdots  &  \vdots   \\
f_{M_R}(z_1) &  \cdots  &  f_{M_R}(z_{M}) \\
g_1(z_1)z_1^{N} &  \cdots  &  g_1(z_M)z_M^{N}  \\
\vdots   &  \vdots  &  \vdots   \\
g_{M_L-1}(z_1)z_1^{N} &  \cdots  &  g_{M_L-1}(z_M)z_M^{N}  \\
g_{M_L}(z_1)z_1^{N} &  \cdots  &  g_{M_L}(z_M)z_M^{N}  \\
\end{array}
\right),~~~
H_{B}^b=\left(
\begin{array}{ccc}
f_1(z_1) &  \cdots  &  f_1(z_{M}) \\
\vdots   &  \vdots  &  \vdots   \\
f_{M_R}(z_1) &  \cdots  &  f_{M_R}(z_{M}) \\
g_1(z_1)z_1^{N} &  \cdots  &  g_1(z_M)z_M^{N}  \\
\vdots   &  \vdots  &  \vdots   \\
g_{M_L-1}(z_1)z_1^{N} &  \cdots  &  g_{M_L-1}(z_M)z_M^{N}  \\
-\delta_tz_1^p &  \cdots  &  -\delta_tz_M^p  \\
\end{array}
\right).
\end{equation}
For $H_B^a$, we have
\begin{equation}
\begin{split}
\det[H_B^a]=&\sum_{i_1\neq i_2 \neq \cdots \neq i_M=1}^{M}(-1)^{\tau(i_1\cdots i_M)}\left[f_1(z_{i_1})\cdots f_{M_R}(z_{i_{M_R}})g_1(z_{i_{M_R+1}})\cdots g_{M_L}(z_{i_M})\left(z_{i_{M_R+1}}z_{i_{M_R+2}}\cdots z_{i_{M}}\right)^{N}\right],
\end{split}
\end{equation}
where $\tau(i_1\cdots i_M)$ is the sorting function of $(i_1\cdots i_M)$. In the thermodynamic limit $N\to\infty$ (or $N\gg1$), the leading term $A_1$ and the second leading term $A_2$ of $\det[H_B^a]$ are
\begin{equation}
\begin{split}
A_1=&\left(z_{M_R+1}z_{M_R+2}\cdots z_{M}\right)^{N}G_a\sum_{i_1 \neq \cdots \neq i_{M_R}=1}^{M_R}(-1)^{\tau(i_1\cdots i_{M_R})}\left[f_1(z_{i_1})\cdots f_{M_R}(z_{i_{M_R}})\right],\\
A_2=&\left(z_{M_R}z_{M_R+2}\cdots z_{M}\right)^{N}G'_a\sum_{i_1 \neq \cdots \neq i_{M_R}=1}^{M_R-1,M_R+1}(-1)^{\tau(i_1\cdots i_{M_R})}\left[f_1(z_{i_1})\cdots f_{M_R}(z_{i_{M_R}})\right],
\end{split}
\end{equation}
with
\begin{equation}
\begin{split}
G_a=&\sum_{i_{M_R+1} \neq \cdots \neq i_M=M_R+1}^{M}(-1)^{\tau(i_{M_R+1}\cdots i_{M})}\left[g_1(z_{i_{M_R+1}})\cdots g_{M_L}(z_{i_M})\right],\\
G'_a=&\sum_{i_{M_R+1} \neq \cdots \neq i_M=M_R,M_R+2}^{M}(-1)^{\tau(i_{M_R+1}\cdots i_{M})}\left[g_1(z_{i_{M_R+1}})\cdots g_{M_L}(z_{i_M})\right].
\end{split}
\end{equation}
The determinant $\det[H_B^b]$ which contains $\delta_t$, can be expanded as
\begin{equation}
\begin{split}
\det[H_B^b]=&-\delta_t\sum_{i_1 \neq \cdots \neq i_M=1}^{M}(-1)^{\tau(i_1\cdots i_M)}\left[f_1(z_{i_1})\cdots f_{M_R}(z_{i_{M_R}})g_1(z_{i_{M_R+1}})\cdots g_{M_L-1}(z_{i_{M-1}})\left(z_{i_{M_R+1}}z_{i_{M_R+2}}\cdots z_{i_{M-1}}\right)^{N}z_{i_{M}}^{p}\right].
\end{split}
\end{equation}
For the relevant case of $N\gg l\gg 1$, the leading term $B_1$ of $\det[H_B^b]$ is
\begin{equation}
\begin{split}
B_1=&-\delta_tz_{M_R+1}^p\left(z_{M_R+2}\cdots z_{M}\right)^{N}G_b\sum_{i_1 \neq \cdots \neq i_{M_R}=1}^{M_R}(-1)^{\tau(i_1\cdots i_{M_R})}\left[f_1(z_{i_1})\cdots f_{M_R}(z_{i_{M_R}})\right],
\end{split}
\end{equation}
where
\begin{equation}
G_b=\sum_{i_{M_R+1} \neq \cdots \neq i_{M-1}=M_R+2}^{M}(-1)^{\tau(i_{M_R+1}\cdots i_{M-1}(M_R+1))}\left[g_1(z_{i_{M_R+1}})\cdots g_{M_L-1}(z_{i_{M-1}})\right].
\end{equation}

For the OBC case with $\delta_t=0$, $\det[H_B] = \det[H_B^a]$. Thus $\det[H_B]= A_1+A_2=0$ in the thermodynamic limit, the condition of generalized Brillouin zone is naturally obtained $|z_{M_R}|=|z_{M_R+1}|$.

For $\delta_t\neq 0$, $\det[H_B]=\det[H_B^a]+\det[H_B^b]=A_1+A_2+B_1$. It is evident that the occurrence of STL requires $|B_1|\gg|A_2|$, i.e.,
\begin{equation}\label{Gcondition}
\delta_t|z_{M_R+1}|^p\gg |z_{M_R}|^N.
\end{equation}
Then $\det[H_B]=A_1+B_1=0$, yielding
\begin{equation}\label{Gzsolu}
z_{M_R+1}=\sqrt[l]{\delta_t\eta}e^{i\theta_m},
\end{equation}
where
\begin{equation}
\eta=\frac{G_b}{G_a},
\end{equation}
and $\theta_m=\frac{2m\pi}{l}$ with $m=1,\cdots,l$. Since $G_a$ and $G_b$ are finite polynomials, $|z_{M_R+1}|^l=|\delta_t\eta|\sim \mathcal{O}(1)$, indicating there are $l$ scale-tailored localized states. Substituting Eq. (\ref{Gzsolu}) into Eq. (\ref{Gcondition}), the condition in Eq. (\ref{Gcondition}) reduces to
\begin{equation}\label{GMTT}
\delta_t > |z_{M_R}|^{l}.
\end{equation}
which governs the existence of $l$ scale-tailored localized states with $|z_{M_R+1}|=\sqrt[l]{\delta_t\eta}$ under the limit $N\gg l\gg1$. Now we relate the above condition to the energy spectra under periodic boundary condition, i.e., the Bloch spectra. For sufficiently large $l$, $|z_{M_R+1}| \rightarrow 1$, and the eigenenergies of these scale-tailored states saturate to the Bloch spectra (i..e, the energy spectra under periodic boundary conditions):
\begin{equation}\label{1}
E = E^{(PBC)} = \sum_{j=1}^{M_R} \frac{t_{jR}}{(e^{ik})^j} + \sum_{j=1}^{M_L} t_{jL} (e^{ik})^j,
\end{equation}
where $k \in [0, 2\pi]$.
Moreover, the condition Eq. (\ref{GMTT}) holds regardless of the magnitude of $\delta_t$ only if $|z_{M_R}|<1$.
For a given $E$ inside the Bloch spectra, there must exist a $z$-solution of unit modulus, which corresponds to the $z$-solutions of the scale-tailored states. Thus, the criterion can be formulated in terms of the Bloch spectra: for any $E \in \sigma_{PBC}$, there exist $M_L + M_R$ solutions $z_i$, sorted by their moduli $\left|z_1\right| \le \left|z_2\right| \le \ldots \le \left|z_{M_L+M_R}\right|$. If these solutions $z_i$ further satisfy $\left|z_{M_R}\right| < \left|z_{M_R+1}\right| = 1$, then $l$ scale-tailored localized states appear upon introducing the additional long-range coupling of arbitrary strength.

\subsection*{(a) The case of $M_R=1$ and $M_L=1$}
In the case of $M_R=1$ and $M_L=1$, the hopping terms are to the nearest neighbors. We denote $r=\sqrt{t_{1R}/t_{1L}}$ and set $t_{1L}=1$ for convenience, thus $\eta=1$. In this simple case, the two $z$-solutions satisfy $|z_1z_2|=r^2$. Our general criterion immediately implies the occurrence of STL when $\delta_t > |z_{1}|^{l}$, and for the scale-tailored states, $|z_{2}|=\sqrt[l]{\delta_t}$. Since $|z_1z_2|=r^2$, this condition reduces to $\delta_t > r^{l}$, which is consistent with the exact solution discussed in Sec. (I). Notably, when $r<1$, the STL occurs for any $\delta_t$. For any finite $\delta_t$, our criterion also determines the condition under which the STL occurs. In the limits of $N\gg l\gg 1$, the $z$-solutions of the Bloch spectra are $|z_1|=r^2$ and $|z_2|=1$ if $r<1$, or $|z_1|=1$ and $|z_2|=r^2$ if $r>1$. Thus, the criterion tells us that the STL occurs when $r<1$.

\subsection*{(b) The case of $M_R=1$ and $M_L=2$}
The Hamiltonian is
\begin{equation}\label{GHb}
\hat{H}=\sum\limits_{n=0}^{N-j-1}\sum\limits_{j=1}^{2}t_{jL}\hat{c}_n^{\dag}\hat{c}_{n+j}+\sum\limits_{n=0}^{N-2}t_{1R}\hat{c}_{n+1}^{\dag}\hat{c}_n +\delta_t \hat{c}_{N-1}^{\dag }\hat{c}_p.
\end{equation}
In Figs. \ref{figSMNew1}(a) and (b), we present its energy spectra in the complex plane for $(N,l)=(100,40)$ and $(N,l)=(100,50)$, respectively. The eigenvalues are categorized into loop-shaped (marked by red or blue circles) and arc-shaped (marked by magenta circles). The Bloch spectra (in green) are included for reference. In Figs. \ref{figSMNew1}(c) and (d), the spatial profiles of the eigenstates corresponding to the loop-shaped eigenvalues are displayed. These eigenstates accumulate towards the $(N-1)$-th site with different localization lengths. In Fig. \ref{figSMNew1}(e), we present the solutions $z_i$ derived from $E^{(PBC)}$, ordered as $|z_1|\leq|z_2|\leq|z_3|$. It is evident that the condition $|z_1|<|z_2|=1$ is satisfied, thus fulfilling our criterion. In Fig. \ref{figSMNew1}(f), we plot the rescaled spatial distributions by the coupling range $l$ of the loop-shaped eigenstates for $(N,l)=(100,40)$ and $(N,l)=(100,50)$. Their perfect overlap confirms that they are indeed scale-tailored localized states.
\begin{figure}[!h]
\includegraphics[width=0.85\textwidth]{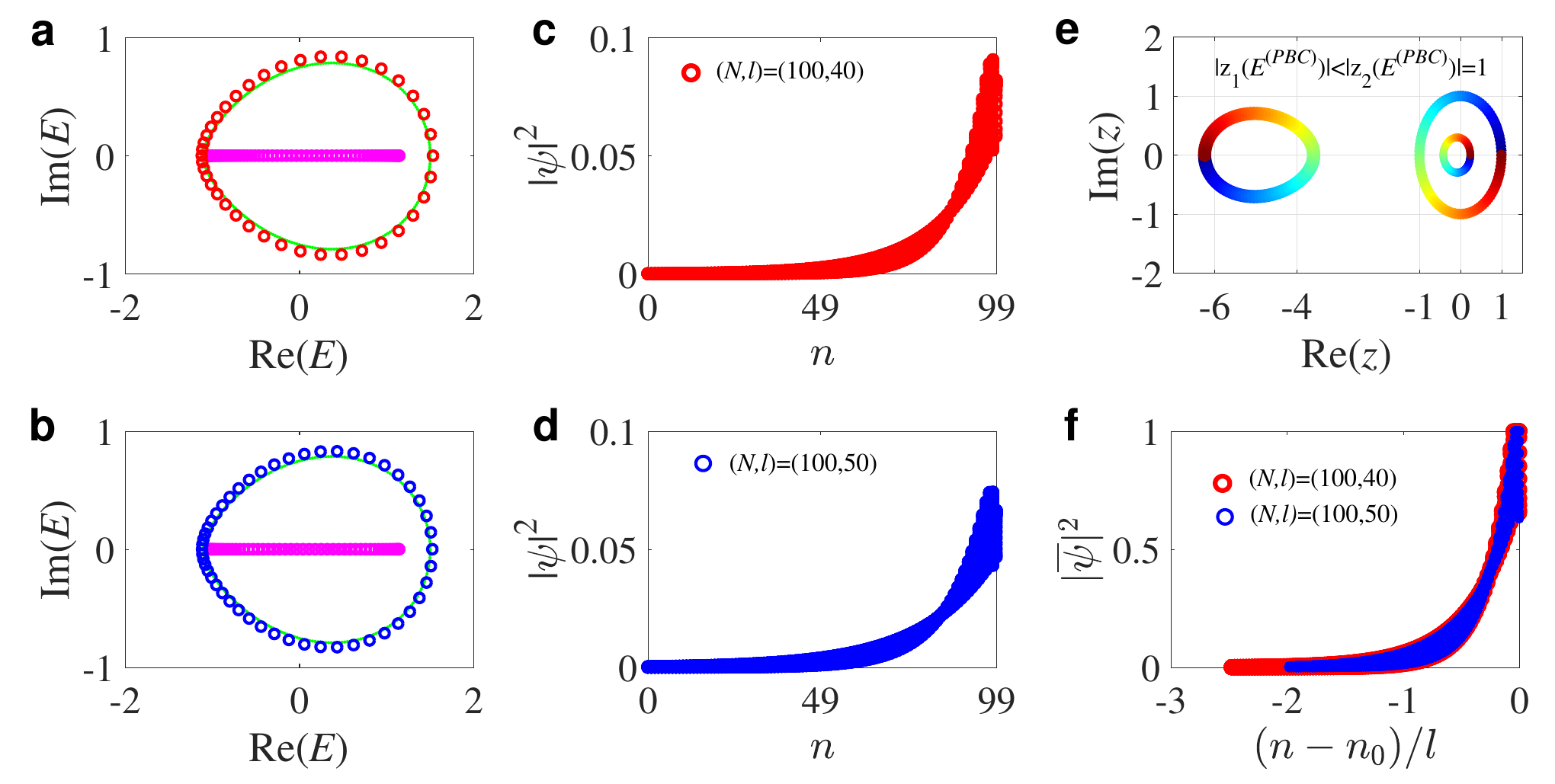}
\caption{Eigenspectra and eigenstates for model Eq. (\ref{GHb}) with paremeters $t_{1L}=1$, $t_{2L}=0.2$, $t_{1R}=0.3$, and $\delta_t=5$. (a)(b) Eigenenergies in the complex plane for $(N,l)=(100,40)$ and $(N,l)=(100,50)$, respectively. The eigenvalues are categorized into loop-shaped (marked by red or blue circles) and arc-shaped (marked by magenta circles). The Bloch spectra are shown in green. (c)(d) Spatial profiles of eigenstates corresponding to the loop-shaped eigenvalues for $(N,l)=(100,40)$ and $(N,l)=(100,50)$, respectively. (e) The $z$-solutions of the Bloch spectra. The data points of the same color represent multiple solutions of $z_i$ corresponding to the same eigenenergy. (f) Rescaled spatial distributions by the coupling range $l$ of the eigenstates in (c)(d).}
\label{figSMNew1}
\end{figure}

\subsection*{(c) The case of $M_R=2$ and $M_L=1$}
The Hamiltonian is
\begin{equation}\label{GHc}
\hat{H}=\sum\limits_{n=0}^{N-2}t_{1L}\hat{c}_n^{\dag}\hat{c}_{n+1}+\sum\limits_{n=0}^{N-j-1}\sum\limits_{j=1}^{2}t_{jR}\hat{c}_{n+j}^{\dag}\hat{c}_n+\delta_t \hat{c}_{N-1}^{\dag }\hat{c}_p.
\end{equation}
In Fig. \ref{figSMNew2}(a) and (b), we present its energy spectra in the complex plane for $(N,l)=(100,40)$ and $(N,l)=(100,50)$, respectively. Similar to the previous case, the eigenvalues are categorized into loop-shaped (marked by red or blue circles) and arc-shaped (marked by magenta circles). Figures \ref{figSMNew2}(c) and (d) display the spatial profiles of the eigenstates corresponding to the loop-shaped spectra for $(N,l)=(100,40)$ and $(N,l)=(100,50)$, respectively. In Fig. \ref{figSMNew2}(e), the $z$-solutions of the Bloch spectra are presented. It is evident that $|z_2|<|z_3|=1$ holds, thus fulfilling our criterion. In Fig. \ref{figSMNew2}(f), we show the rescaled spatial distributions for $(N,l)=(100,40)$ and $(N,l)=(100,50)$, confirming their STL nature.
\begin{figure}[!h]
\includegraphics[width=0.85\textwidth]{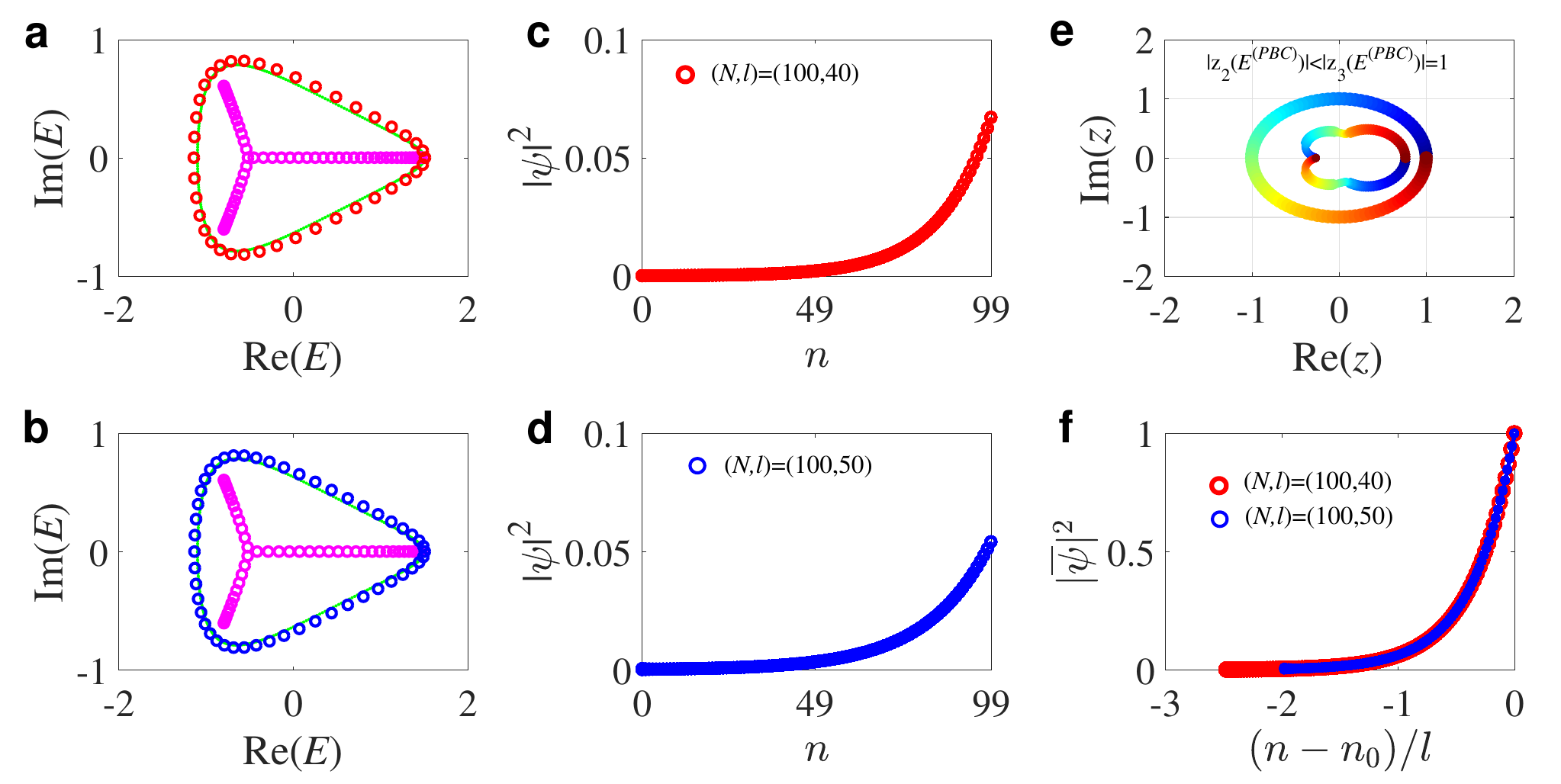}
\caption{Eigenspectra and eigenstates for model Eq. (\ref{GHc}) with paremeters $t_{1L}=1$, $t_{1R}=0.3$, $t_{2R}=0.2$, and $\delta_t=4$. (a)(b) Eigenenergies in the complex plane for $(N,l)=(100,40)$ and $(N,l)=(100,50)$, respectively. The spectra are categorized into loop-shaped (marked by red or blue circles) and arc-shaped (marked by magenta circles). The Bloch spectra are shown in green. (c)(d) Spatial profiles of eigenstates corresponding to the loop-shaped eigenvalues for $(N,l)=(100,40)$ and $(N,l)=(100,50)$, respectively. (e) The $z$-solutions of the Bloch spectra. (f) Rescaled spatial distributions by the coupling range $l$ of the eigenstates in (c)(d).}
\label{figSMNew2}
\end{figure}

\section*{(III) Scale-tailored localized states in the vicinity of impurities}
\begin{figure}[!h]
\includegraphics[width=0.85\textwidth]{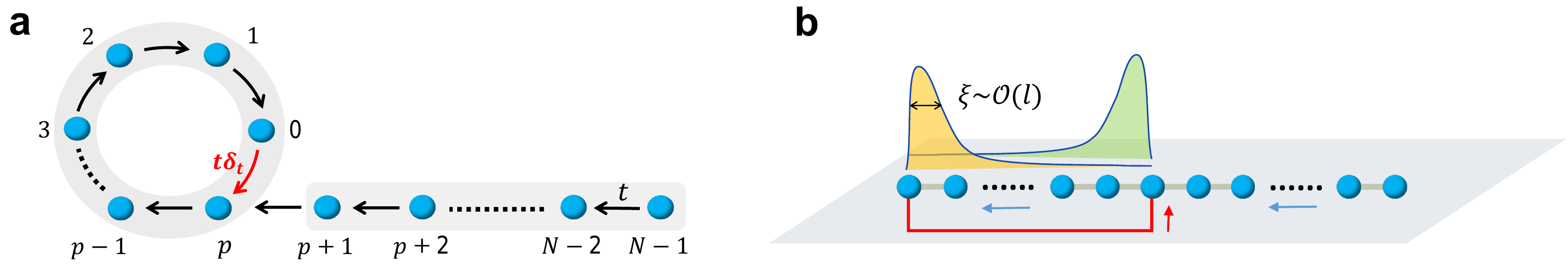}
\caption{(a) Schematics of the lattice model with rolled boundary condition described by Eq. (\ref{SH2}). (b) Sketch of the localization behavior for the (inverse) scale-tailored localized states in (a): the scale-tailored localized states are localized at the boundary with localization length $\xi\propto l$, while the inverse scale-tailored localized states are localized near the impurity with $\xi\propto l$.}
\label{figSM3}
\end{figure}
In addition to being localized at the boundary, the scale-tailored localized states induced by long-range asymmetric couplings can also reside in the vicinity of impurities. Here, we take the example of the unidirectional hopping model with rolled boundary conditions as sketched in Fig. \ref{figSM3}(a). The Hamiltonian is
\begin{equation}\label{SH2}
\hat{H}=\sum\limits_{i=0}^{N-2}t\hat{c}_{i}^{\dag }\hat{c}_{i+1}+t\delta_t\hat{c}_{p}^{\dag }\hat{c}_{0},
\end{equation}
where $N$ is the number of lattice sites, $t$ is the hopping amplitude, and $\delta_t$ is the coupling between site $p$ and the first site. This rolled boundary condition forms a closed loop of length $l=p+1$. Similarly to the previous section, we have the bulk equation
\begin{equation}\label{SH1BE}
-E\psi _{n}+t\psi _{n+1}=0,
\end{equation}
with $n=0,1,\cdots,p-1,p+1,\cdots,N-2$. The boundary equations are
\begin{equation}\label{SH2BdE}
\begin{split}
-E\psi _{N-1}&=0,\\
t\delta_t\psi _{0}-E\psi _{p}&=0.
\end{split}
\end{equation}
Due to the discontinuity of the bulk equations at $n=p$, we take the ansatz wave function $\Psi$ as
\begin{equation}\label{SH2wv}
(\psi _{0},\psi _{1},\cdots,\psi _{p},\psi _{p+1},\psi _{p+2},\cdots,\psi _{N-1})^{T}=(1,z,\cdots,z^{p},\phi,z\phi,\cdots,z^{N-1-l}\phi)^{T}.
\end{equation}
Here $\phi$ is an undetermined parameter to distinguish the components of the wave function at rolled sites and the remaining sites. Inserting Eq. (\ref{SH2wv}) into bulk equations Eq. (\ref{SH1BE}), the eigenvalue can be obtained as
\begin{equation}\label{SH1E}
E=tz.
\end{equation}
Further substituting Eq. (\ref{SH2wv}) into Eq. (\ref{SH2BdE}) and combining with Eq. (\ref{SH1E}), we have
\begin{equation}\label{SZphieq}
\begin{split}
z^{N-l}\phi&=0,\\
z^{l}-\delta_t-\phi&=0.
\end{split}
\end{equation}
For periodic ($\delta_t=1$ and $p=N-1$) or open ($\delta_t=0$) boundary conditions, the model is the same as Eq. (1) in the main text, which we omit here. For the more general cases of $\delta_t\neq 0$ and $p\neq N-1$, the solutions can be classified into two types. The first type comprises $(N-l)$ degenerate solutions ($(N-l)$-th order EP):
\begin{equation}
\begin{split}
E&=0,\\
\Psi&=(1,0,\cdots,0,\delta_t,0,\cdots,0)^{T},
\end{split}
\end{equation}
with $z=0$ and $\phi=\delta_t$. It is worth noting that these states are localized at both the first site and impurity ($p$-th) site. The second type corresponds to solutions $z^{(m)}=\sqrt[l]{\delta_t}e^{i\theta_m}$ with $\theta_m=\frac{2m\pi}{l}~(m=1,2,\cdots,l)$ and $\phi=0$:
\begin{equation}
\begin{split}
E_m&=t\sqrt[l]{\delta_t}e^{i\theta_m},\\
\Psi^{(m)}&=(1,\sqrt[l]{\delta_t}e^{i\theta_m},\cdots,(\sqrt[l]{\delta_t}e^{i\theta_m})^{p},0,0,\cdots,0)^{T}.
\end{split}
\end{equation}
These $l$ eigenenergies are evenly distributed on a circle of radius $t\sqrt[l]{\delta_t}$. These eigenstates are localized exclusively on the rolled lattice sites, with localization center at $n_0=0$ (for $|\delta_t|<1$) or $n_0=p$ (for $|\delta_t|>1$). The localization length $\xi$ of these $l$ eigenstates is
\begin{equation}\label{SH1xi}
\begin{split}
\xi&=
\left\{
  \begin{array}{ll}
    -\frac{l}{\log|\delta_t|},~~~\hbox{$|\delta_t|<1$,} \\
    \frac{l}{\log|\delta_t|},~~~~~\hbox{$|\delta_t|>1$.}
  \end{array}
\right.
\end{split}
\end{equation}
Depending on whether the additional coupling is weaker or stronger than the unidirectional hopping in the bulk, these $l$ eigenstates accumulate either at the left boundary or the vicinity of impurity with the localization length proportional to the coupling range $l$. That is, they are scale-tailored localized or inverse scale-tailored localized states. In the complex energy plane, the degenerate states (exceptional point) and the (inverse) scale-tailored localized states are separated.

\section*{(IV) Scale-tailored localized states induced by multiple long-range asymmetric couplings}
\begin{figure}[b]
\includegraphics[width=0.7\textwidth]{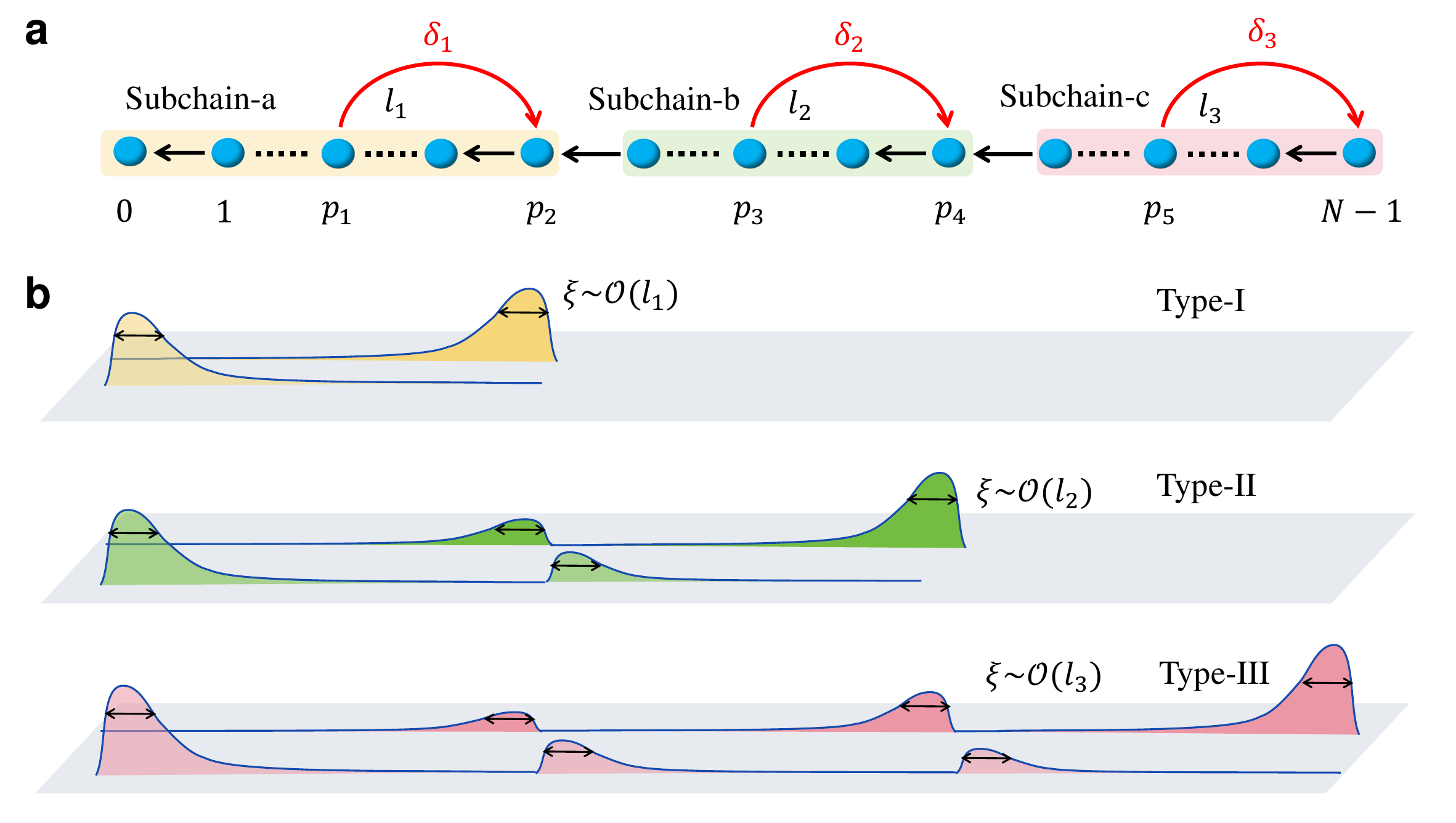}
\caption{(a) Sketch of the unidirectional hopping model with three asymmetric long-range couplings. (b) Sketch of three types of scale-tailored localized states: Type-I reside exclusively on subchain-a, with localization length $\xi\propto l_1$; Type-II reside at subchain-a and subchain-b, with $\xi\propto l_2$; Type-III resides at all three subchains, with $\xi\propto l_3$.}
\label{figSMNew3}
\end{figure}
In this section, we examine the case of multiple long-range asymmetric couplings and demonstrate the occurrence of STL. As an example, we consider the unidirectional hopping model featuring three asymmetric long-range couplings, as illustrated in Fig. \ref{figSMNew3}(a). This model can be exactly solved. The Hamiltonian is
\begin{equation}\label{GH2Im}
\hat{H}=\sum\limits_{n=0}^{N-2}t\hat{c}_n^{\dag}\hat{c}_{n+1}+\delta_{1} \hat{c}_{p_2}^{\dag }\hat{c}_{p_1}+\delta_{2} \hat{c}_{p_4}^{\dag }\hat{c}_{p_3}+\delta_{3} \hat{c}_{N-1}^{\dag }\hat{c}_{p_5}.
\end{equation}
Here, $N$ represents the length of the lattice, $t$ denotes the hopping amplitude and we set $t=1$ for convience. The three asymmetric couplings have strength $\delta_{1}$, $\delta_{2}$, and $\delta_{3}$, extending respectively from the $p_1$-th to the $p_2$-th site, the $p_3$-th to the $p_4$-th site, and from the $p_5$-th site to the last site. We take $0<p_1<p_2<p_3<p_4<p_5<(N-1)$ and the coupling ranges are $l_1=p_2-p_1+1$, $l_2=p_4-p_3+1$ and $l_3=N-p_5$. The eigenvalue equation for model (\ref{GH2Im}) is
\begin{equation}
\hat{H}|\Psi\rangle=E|\Psi\rangle,
\end{equation}
with $|\Psi\rangle=\sum_{n}\psi _{n}c_{n}^{\dag}|0\rangle$. As the chain is divided into three parts (labeled as subchain-a/b/c), we take an appropriate wave function as
\begin{equation}\label{GH2Imwvg}
\Psi(z)=(1,z,\cdots,z^{p_2},z^{p_2+1}\phi_a,z^{p_2+2}\phi_a,\cdots,z^{p_4}\phi_a,z^{p_4+1}\phi_b,z^{p_4+2}\phi_b,\cdots,z^{N-1}\phi_b)^{T},
\end{equation}
where $\phi_a$ and $\phi_b$ are parameters to be determined. For bulk lattice sites,
\begin{equation}\label{GH2ImBE}
-E\psi _{n}+\psi _{n+1}=0,
\end{equation}
with $n=0,1,\cdots,p_2-1,p_2+1,\cdots,p_4-1,p_4+1,\cdots,N-2$. For impurity sites and boundary sites, we have
\begin{equation}\label{GH2ImBdE}
\begin{split}
\delta_{1} \psi _{p_1}-E\psi _{p_2}+\psi _{p_2+1}&=0,\\
\delta_{2} \psi _{p_3}-E\psi _{p_4}+\psi _{p_4+1}&=0,\\
\delta_{3} \psi _{p_5}-E\psi _{N-1}&=0.\\
\end{split}
\end{equation}
By substituting the ansatz Eq. (\ref{GH2Imwvg}) into Eqs. (\ref{GH2ImBE}), we obtain the eigenvalues
\begin{equation}\label{GH2ImE}
E=z.
\end{equation}
Further inserting Eq. (\ref{GH2Imwvg}) and Eqs. (\ref{GH2ImE}) into Eqs. (\ref{GH2ImBdE}), we obtain
\begin{equation}
\begin{split}
z^{p_1}\left(\delta_{1}-z^{l_1}+z^{l_1}\phi_a\right)&=0,\\
z^{p_3}\left(\delta_{2}\phi_a-z^{l_2}\phi_a+z^{l_2}\phi_b\right)&=0,\\
z^{p_5}\phi_b\left(\delta_{3}-z^{l_3}\right)&=0.
\end{split}
\end{equation}
Combining the above three equations yields the following two classes of solutions:

(1) Class-I: There are $(N-l_1-l_2-l_3)$-fold degenerate solutions (exceptional points) $z=0$.
\begin{equation}
\begin{split}
E&=0,\\
\Psi&=(1,0,\cdots ,0)^{T}.
\end{split}
\end{equation}
These states are localized exclusively at the first site.

(2) Class-II: The remaining $(l_1+l_2+l_3)$ solutions correspond to scale-tailored localized states and can be further classified into three types:

(i) Type-I corresponds to solutions $z^{(m)}=\sqrt[l_1]{\delta_1}e^{i\theta_m}$ with $\theta_m=\frac{2m\pi}{l_1}~(m=1,\cdots,l_1)$, $\phi_a=0$ and $\phi_b=0$. The eigenvalues and eigenfunctions take:
\begin{equation}
\begin{split}
E_m&=\sqrt[l_1]{\delta_1}e^{i\theta_m},\\
\Psi^{(m)}&=\left(1,\sqrt[l_1]{\delta_1}e^{i\theta_m},\cdots ,\left(\sqrt[l_1]{\delta_1}e^{i\theta_m}\right)^{p_2},0,\cdots,0\right)^{T}.
\end{split}
\end{equation}
These $l_1$ eigenenergies are evenly distributed on a circle of radius $\sqrt[l_1]{\delta_1}$. As depicted in the upper panel of Fig. \ref{figSMNew3}(b), these eigenstates reside exclusively on subchain-a, centered at $n_0=0$ (for $|\delta_1|<1$) or $n_0=p_2$ (for $|\delta_1|>1$), with localization length
\begin{equation}\label{GH2xi}
\begin{split}
\xi&=
\left\{
  \begin{array}{ll}
    -\frac{l_1}{\log|\delta_1|},~~~\hbox{$|\delta_1|<1$,} \\
    \frac{l_1}{\log|\delta_1|},~~~~~\hbox{$|\delta_1|>1$.}
  \end{array}
\right.
\end{split}
\end{equation}
Thus, they are scale-tailored localized states due to $\xi\propto l_1$.

(ii) Type-II has solutions $z^{(m)}=\sqrt[l_2]{\delta_2}e^{i\theta_m}$ with $\theta_m=\frac{2m\pi}{l_2}~(m=1,\cdots,l_2)$, along with $\phi_b=0$ and $\phi_a=1-\frac{\delta_1}{z^{l_1}}=1-\delta_1/(\sqrt[l_2]{\delta_2}e^{i\theta_m})^{l_1}$.
The eigenvalues and eigenfunctions are
\begin{equation}
\begin{split}
E_m&=\sqrt[l_2]{\delta_2}e^{i\theta_m},\\
\Psi^{(m)}&=\left(1,\cdots,\left(\sqrt[l_2]{\delta_2}e^{i\theta_m}\right)^{p_2},\left(\sqrt[l_2]{\delta_2}e^{i\theta_m}\right)^{p_2+1}\phi_a,\cdots,\left(\sqrt[l_2]{\delta_2}e^{i\theta_m}\right)^{p_4}\phi_a,0,\cdots,0\right)^{T}.
\end{split}
\end{equation}
These $l_2$ eigenenergies are evenly distributed on a circle of radius $\sqrt[l_2]{\delta_2}$. As illustrated in the middle panel of Fig. \ref{figSMNew3}(b), the eigenstates are solely distributed on subchain-a and subchain-b, split into two segments. Depending on the magnitude of $|\delta_2|$, the localization centers  of these two segments are $n_0^{(1)}=p_2$ and $n_0^{(2)}=p_4$ for $|\delta_2|>1$, and $n_0^{(1)}=0$ and $n_0^{(2)}=p_2+1$ for $|\delta_2|<1$. Both segments exhibit the same localization length:
\begin{equation}\label{GH2xi}
\begin{split}
\xi&=
\left\{
  \begin{array}{ll}
    -\frac{l_2}{\log|\delta_2|},~~~\hbox{$|\delta_2|<1$,} \\
    \frac{l_2}{\log|\delta_2|},~~~~~\hbox{$|\delta_2|>1$.}
  \end{array}
\right.
\end{split}
\end{equation}
These $l_2$ states are also scale-tailored localized eigenstates.

\begin{figure}[htb]
\includegraphics[width=0.93\textwidth]{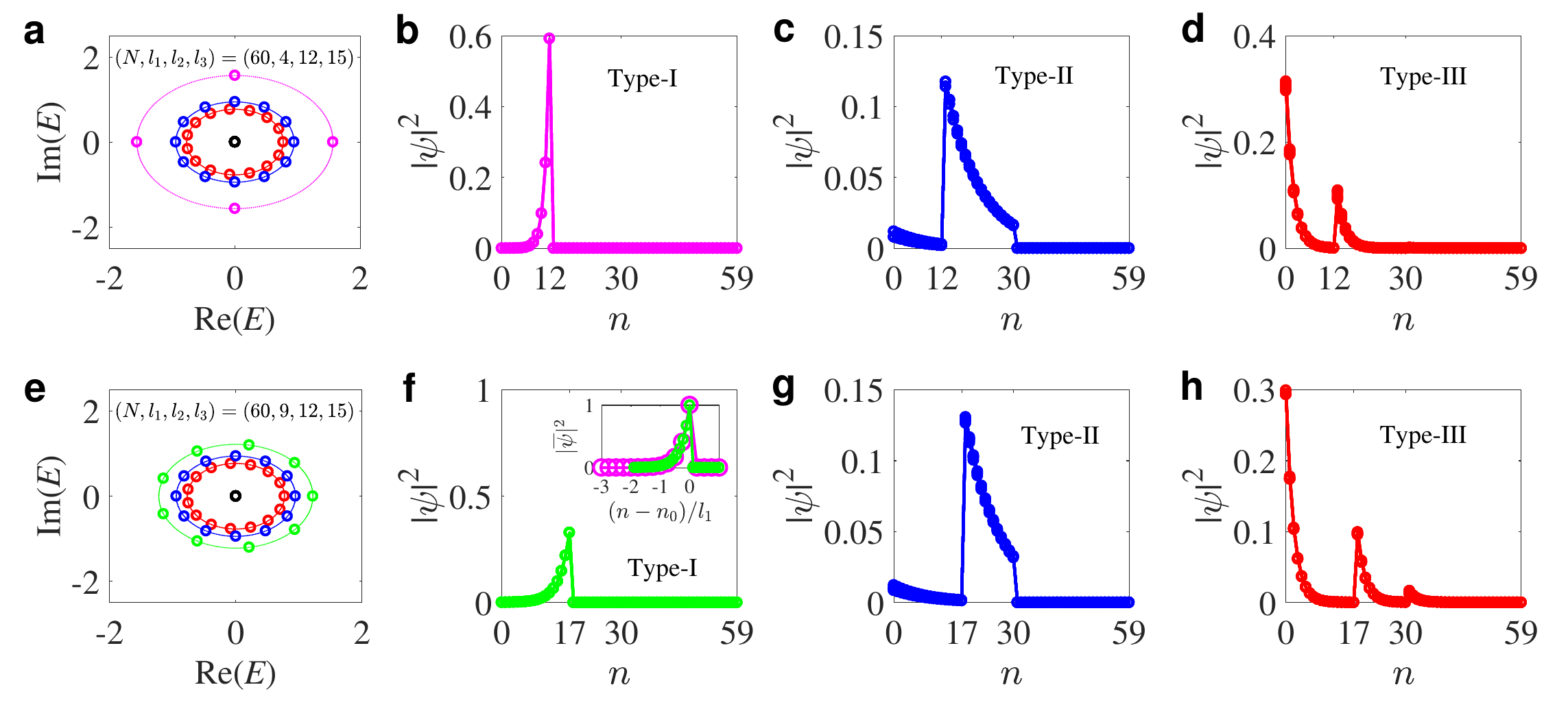}
\caption{Eigenspectra and eigenstates of model (\ref{GH2Im}), with three asymmetric long-range couplings. (a) Eigenspectra for $(N,l_1,l_2,l_3)=(60,4,12,15)$ and $p_2=12$, where circles/dots represent numerical/analytical results. (b-d) The three types of scale-tailed localized states corresponding to the eigenspectra in (a). (e) Eigenspectra for $(N,l_1,l_2,l_3)=(60,9,12,15)$ and $p_2=17$, where circles/dots represent numerical/analytical results. (f-h) The three types of scale-tailed localized states corresponding to the eigenspectra in (e). Inset of (f): Perfect overlapping of the rescaled spatial profiles by coupling range $l_1$ of the Type-I eigenstates. Other parameters: $\delta_1=6$, $\delta_2=0.5$, $\delta_3=0.02$, $p_1=9$, $p_3=19$, $p_4=30$, $p_5=45$.}
\label{figSMNew4}
\end{figure}

(iii) Type-III corresponds to solutions  $z^{(m)}=\sqrt[l_3]{\delta_3}e^{i\theta_m}$ where $\theta_m=\frac{2m\pi}{l_3}~(m=1,\cdots,l_3)$, and $\phi_a=1-\frac{\delta_1}{z^{l_1}}=1-\delta_1/(\sqrt[l_3]{\delta_3}e^{i\theta_m})^{l_1}$, and $\phi_b=(1-\frac{\delta_2}{z^{l_2}})\phi_a$. The eigenvalues and eigenfunctions are:
\begin{equation}
\begin{split}
E_m&=\sqrt[l_3]{\delta_3}e^{i\theta_m},\\
\Psi^{(m)}&=\left[1,\cdots,\left(\sqrt[l_3]{\delta_3}e^{i\theta_m}\right)^{p_2},\left(\sqrt[l_3]{\delta_3}e^{i\theta_m}\right)^{p_2+1}\phi_a,\cdots,\left(\sqrt[l_3]{\delta_3}e^{i\theta_m}\right)^{p_4}\phi_a,\left(\sqrt[l_3]{\delta_3}e^{i\theta_m}\right)^{p_4+1}\phi_b,,\cdots,\left(\sqrt[l_3]{\delta_3}e^{i\theta_m}\right)^{N-1}\phi_b\right]^{T}.
\end{split}
\end{equation}
These $l_3$ eigenenergies are evenly distributed on a circle of radius $\sqrt[l_3]{\delta_3}$. As depicted in the lower panel of Fig. \ref{figSMNew3}(b), these eigenstates are divided into three segments. Depending on $|\delta_3|$, the localization centers for these three segments are  $n_0^{(1)}=p_2$, $n_0^{(2)}=p_4$ and $n_0^{(2)}=N-1$ for $|\delta_3|>1$, and $n_0^{(1)}=0$, $n_0^{(2)}=p_2+1$ and $n_0^{(3)}=p_4+1$ for $|\delta_3|<1$. All segments exhibit the same localization length:
\begin{equation}\label{GH2xi}
\begin{split}
\xi&=
\left\{
  \begin{array}{ll}
    -\frac{l_3}{\log|\delta_3|},~~~\hbox{$|\delta_3|<1$,} \\
    \frac{l_3}{\log|\delta_3|},~~~~~\hbox{$|\delta_3|>1$.}
  \end{array}
\right.
\end{split}
\end{equation}
Therefore, these $l_3$ states are scale-tailored localized states.

In Fig. \ref{figSMNew4}, we present eigensolutions of the unidirectional hopping model with three asymmetric long-range couplings, comparing two sets of eigensolutions with different $l_1$. Apart from common parameters like $(\delta_1,\delta_2,\delta_3)=(6,0.5,0.02)$, $(N,l_2,l_3)=(40,12,15)$, $(p_1,p_3,p_4,p_5)=(9,19,30,45)$, parameters of the subchain-a in Figs. \ref{figSMNew4}(a)-(d) are $l_1=40$ and $p_2=12$, while Figs. \ref{figSMNew4}(e)-(h) corresponds to $l_1=9$ and $p_2=17$. As illustrated in Figs. \ref{figSMNew4}(a)(e), apart from the eigenvalues that are degenerate at $E=0$, the remaining eigenvalues are distributed on three large circles, marked by pink, blue, and red symbols, corresponding respectively to scale-tailored localized eigenstates of Type-I, Type-II, and Type-III, as depicted in Figs. \ref{figSMNew4}(b)(f), Figs. \ref{figSMNew4}(c)(g), and Figs. \ref{figSMNew4}(d)(h). The Type-I eigenstates reside exclusively on subchain-a, Type-II eigenstates are distributed exclusively between subchain-a and subchain-b, and the type-III eigenstates are divided into three segments distributed among subchain-a, subchain-b, and subchain-c. The inset in Fig.  \ref{figSMNew4}(f) demonstrates the exact overlap of rescaled spatial distributions by the coupling range $l_1$ of Type-I eigenstates shown in Fig. \ref{figSMNew4}(b) and the main image in Fig. \ref{figSMNew4}(f), which indicates that Type-I eigenstates are  tailored localized states with $\xi\propto l_1$.

\section*{(V) The STL in the 2D unidirectional hopping model}

\begin{figure}[bth]
\includegraphics[width=.75\textwidth]{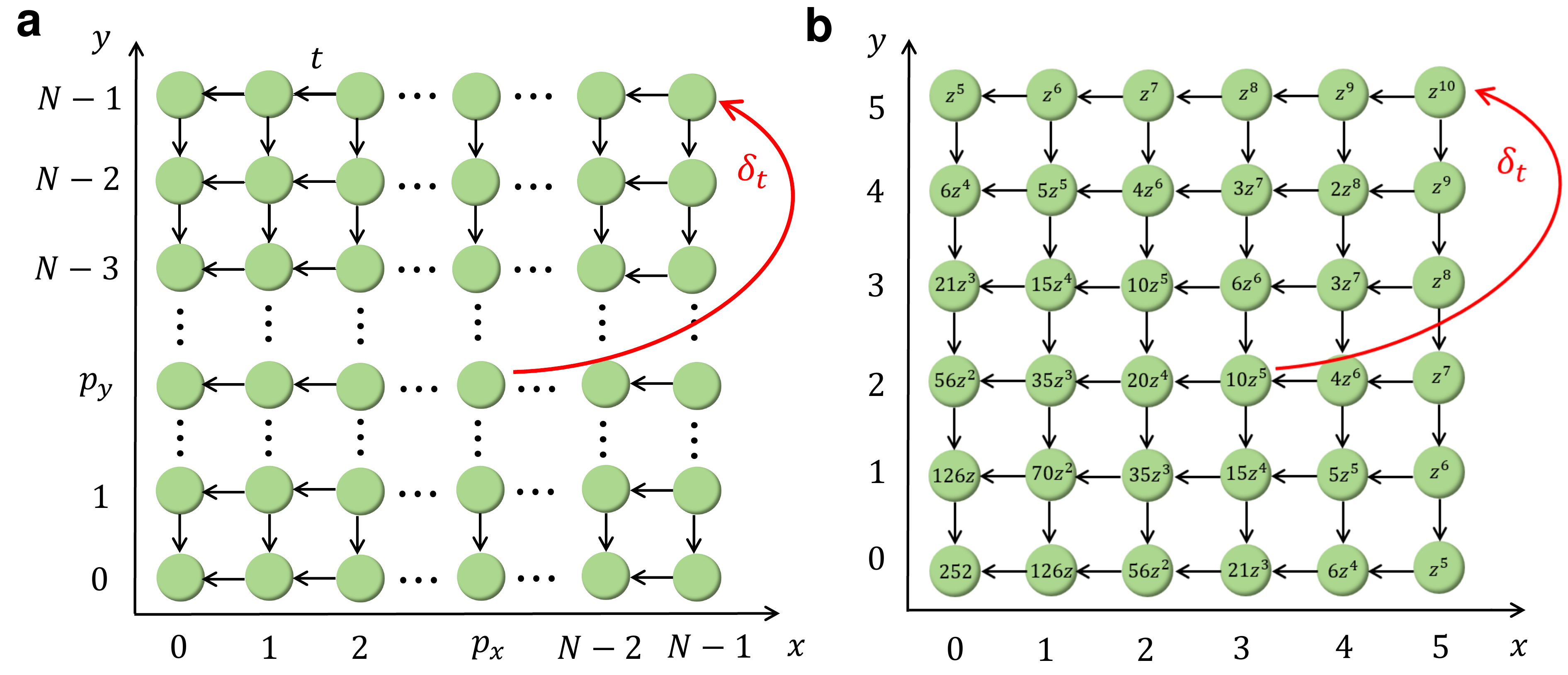}
\caption{ (a) Sketch of the 2D unidirectional hopping model with an additional asymmetry coupling (red arrow). The model is analytically solvable. (b) The coefficients from exact solutions yield an elegant Pascal's Triangle on a $6\times 6$ lattice.}
\label{figSMNew5}
\end{figure}
The spatial dimensionality plays an important role on the localization effect. In this section, we explore the STL in 2D and study an exactly solvable model as depicted in Fig. \ref{figSMNew5}(a). The Hamiltonian is
\begin{equation}
\hat{H}_{2D}=\sum\limits_{j=0}^{N-1}\sum\limits_{i=0}^{N-2}t\hat{c}_{i,j}^{\dag}\hat{c}_{i+1,j}+\sum\limits_{i=0}^{N-1}\sum\limits_{j=0}^{N-2}t\hat{c}_{i,j}^{\dag}\hat{c}_{i,j+1}+\delta _{t}\hat{c}_{N-1,N-1}^{\dag}\hat{c}_{p_{x},p_{y}},
\end{equation}
where $N$ is the number of lattice sites along the $x/y$ direction. This model features unidirectional hoppings along both the $x$ and $y$ directions with strength $t$, and an additional long-range coupling from the $(p_x,p_y)$-th to the last site with strength $\delta_t$. The hopping ranges along the two directions are  $l_x=N-p_x$ and $l_y=N-p_y$, respectively. The egenvalue equation $H\Psi =E\Psi $ consists of the bulk equations
\begin{equation}\label{GH2DE}
-E\psi _{i,j}+\psi _{i+1,j}+\psi _{i,j+1}=0,
\end{equation}
with $i=0,\cdots,N-2$, and $j=0,\cdots ,N-2$, and the boundary equations
\begin{equation}\label{GH2DdE1}
-E\psi _{N-1,j}+\psi _{N-1,j+1}=0\text{
\ \ }(j=0,\cdots ,N-2),
\end{equation}
\begin{equation}\label{GH2DdE2}
-E\psi _{i,N-1}+\psi _{i+1,N-1}=0\text{
\ \ }(i=0,\cdots ,N-2),
\end{equation}
\begin{equation}\label{GH2DdE3}
-E\psi _{N-1,N-1}+\delta _{t}\psi _{p_{x},p_{y}}=0.
\end{equation}
Due to the translational invariance of the bulk equations, we set an appropriate ansatz of wave function $\Psi$ as
\begin{equation}\label{GH2DWvs}
\Psi  =(\psi _{0,0},\psi _{1,0},\cdots ,\psi _{N-1,0},\psi _{0,1},\psi _{1,1},\cdots ,\psi
_{N-1,1},\cdots ,\psi _{0,N-1},\psi _{1,N-1},\cdots ,\psi _{N-1,N-1})^{T},
\end{equation}
where
\begin{equation}\label{GH2DWv}
\psi _{ij}=s_{i,j}z^{i+j},
\end{equation}
with $i=0,\cdots ,N-1$, $j=0,\cdots ,N-1$, and $s_{i,j}$ is a coefficient to be determined. Here we assume $s_{N-1,N-1}=1$ as an initial value.

By inserting Eq. (\ref{GH2DWv}) into Eqs. (\ref{GH2DE}), we obtain
\begin{equation}\label{GH2DEs}
E=\frac{s_{i+1,j}+s_{i,j+1}}{s_{i,j}}z
\end{equation}
with $i=0,\cdots ,N-2$, $j=0,\cdots ,N-2$. Note that $E$ refers to the eigenvalues of this system and is independent of $i$ and $j$, we have
\begin{equation}\label{GH2DEE}
E=z,
\end{equation}
and
\begin{equation}\label{GH2DSS}
\frac{s_{i+1,j}+s_{i,j+1}}{s_{i,j}}=1
\end{equation}
with $i=0,\cdots ,N-2$, $j=0,\cdots ,N-2$. By inserting Eq. (\ref{GH2DWv}) and Eq. (\ref{GH2DEE}) into the boundary equations (\ref{GH2DdE1},\ref{GH2DdE2}) and considering $s_{N-1,N-1}=1$, we have
\begin{equation}\label{GH2DSS1}
s_{N-1,j}=s_{i,N-1}=1,
\end{equation}
where $i=0,\cdots ,N-1$, $j=0,\cdots ,N-1$. Substituting Eq. (\ref{GH2DSS1}) into Eq. (\ref{GH2DSS}) yields the solutions
\begin{equation}\label{GH2DSSk}
s_{i,j}=C_{2N-i-j-2}^{N-i-1},
\end{equation}
where $i=0,\cdots ,N-2$, $j=0,\cdots ,N-2$, and $C_n^m=\frac{n!}{m!(n-m)!}$. These coefficients form Pascal's Triangle, and for a $6\times 6$ lattice, they are illustrated in Fig. \ref{figSMNew5}(b). Inserting Eq. (\ref{GH2DEE}) into Eq. (\ref{GH2DdE3}), we obtain
\begin{equation}
z^{p_x+p_y}\left(\delta _{t}s_{p_x,p_y}-z^{l}\right)=0.
\end{equation}
Our focus is on the scale-tailored localized states governed by $\delta_t s_{p_x,p_y} - z^{l}$. There exist $l$ such non-degenerate states:
\begin{equation}
z^{(m)}=\sqrt[l]{\delta_ts_{p_x,p_y}}e^{i\theta_m},
\end{equation}
where
\begin{equation}
\begin{split}
s_{p_x,p_y}&=\left\{
\begin{aligned}
& 1, \text{~~~~~for~~}  l_x=1 \text{~~or~~} l_y=1;\\
& C_{l-1}^{l_x-1}, \text{~~~~~for~~} l_x\neq1 \text{~~and~~} l_y\neq1.
\end{aligned}
\right.
\end{split}
\end{equation}
with $\theta_m=\frac{2m\pi}{l},~(m=1,2,\cdots,m)$. The localization length relates to the hopping range as
\begin{eqnarray}
\xi\sim\ \frac{l}{\log{\left|\delta_ts_{p_x,p_y}\right|}}.
\end{eqnarray}
It is evident that localization length still highly depends on the coupling ranges in both directions, yet in intriguing ways, making these states distinct from the normal skin modes. In higher dimensions, there exist infinite spatial directions of the additional long-range coupling, which compicates the STL in 2D.
\begin{figure}[hbt]
\includegraphics[width=.8\textwidth]{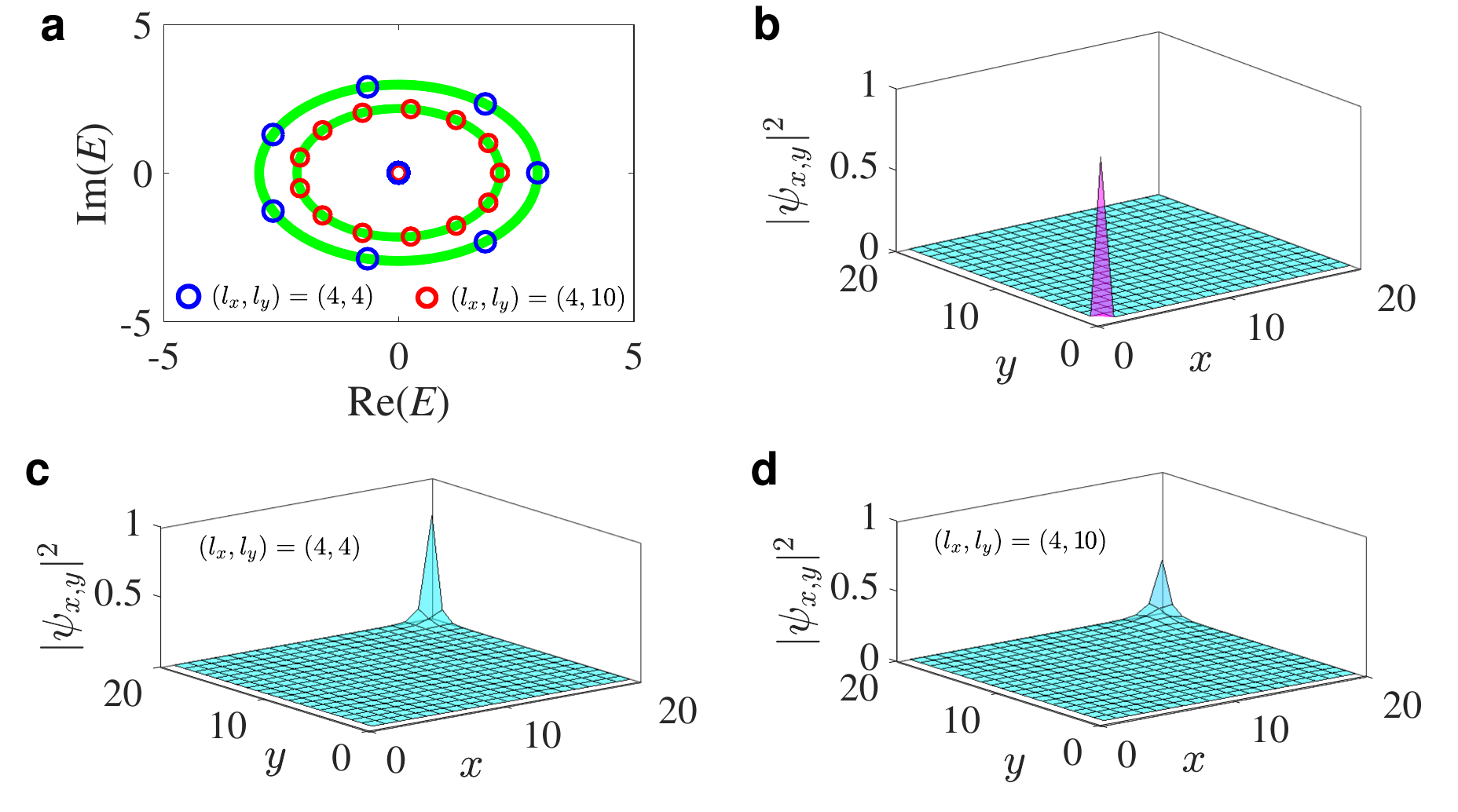}
\caption{Eigenspectra and eigenstates for the 2D unidirectional hopping model on a $20\times20$ lattice. (a) Energy spectra for $(l_x,l_y)=(4,4)$ and $(l_x,l_y)=(4,10)$ are marked with blue and red symbols, respectively. The spectra are divided into two sectors: the first type is $E=0$ (highly degenerate); the second type is evenly distributed on a circle of radius $\sqrt[l]{\delta_ts_{p_xp_y}}$ (represented by green lines). (b) The spatial profile of the first-type eigenstates. (c)-(d) The spatial profile of the second-type eigenstates for $(l_x,l_y)=(4,4)$ and $(l_x,l_y)=(4,10)$, respectively. Other parameters are: $t=1,\delta_t=100$.}
\label{figSMNew6}
\end{figure}

In Fig. \ref{figSMNew6}(a), we display the energy spectra for $\left(l_x,l_y\right)=\left(4,4\right)$ (in blue) and $\left(l_x,l_y\right)=\left(4,10\right)$ (in red) of the 2D system on a $20\times20$ lattice. The energy spectra are divided into two sectors: the first type corresponds to $E=0$, while the second type consists of eigenvalues evenly distributed on a circle of radius $\sqrt[l]{\delta_t s_{ij}}$. The eigenstates of the first type localize exclusively at the $\left(0,0\right)$-th site, as shown in Fig. \ref{figSMNew6}(b). Figures \ref{figSMNew6}(c) and \ref{figSMNew6}(d) depict the eigenstates of the second type for $\left(l_x,l_y\right)=\left(4,4\right)$ and $\left(l_x,l_y\right)=\left(4,10\right)$, respectively. Their localization lengths depend on the coupling ranges along both the $x$ and $y$ directions, governed by our exact results.

\section*{(VI) The STL in the Bose-Hubbard model}

In this section, we explore the impact of many-body interactions on STL using the paradigmatic Bose-Hubbard model. The Hamiltonian is
\begin{equation}
\hat{H}=\sum\limits_{j=1}^{L-1}t\hat{b}_{j}^{\dag}\hat{b}_{j+1}+\delta _{t}\hat{b}_{L}^{\dag}\hat{b}_{p}+\sum\limits_{j=1}^{L}\frac{U}{2}\hat{n}_j(\hat{n}_j-1),
\end{equation}
where $L$ denotes the total number of lattice sites. Here, $\hat{b}_{j}^{\dag}$ and $\hat{b}_{j}$ are bosonic creation and annihilation operators at the $j$-th site, respectively. The operator $\hat{n}_j$ is the particle number operator on the $j$-th site, and $U$ is the strength of on-site interaction. $t$ is the magnitude of unidirectional hopping and $\delta_t$ is the strength of the asymmetric long-range coupling, with coupling range $l=L-p+1$. In the following, we focus on the case of two interacting bosons $N=2$.

We employ exact diagonalization to obtain the eigenspectra and eigenstates. Figs. \ref{figSMNew7}(a)-(d) illustrates the energy spectra for four different interaction strengths: $U=0$, $U=0.5$, $U=1$, and $U=5$, respectively. The red and blue symbols correspond to the cases where $l=16$ and $l=8$. To validate the occurrence of STL, in Figs. \ref{figSMNew7}(e)-(h), we show the spatial distributions of the particle number ${\bar{n}}_j$ for the ground state corresponding to the colored circles in Figs. \ref{figSMNew7}(a)-(d), respectively. (For non-Hermitian systems, the ground state refers to the state with the smallest real part of eigenvalues.) It is evident that for small $U$, the spatial profile decays exponentially to the left. Upon scaling, the profiles for $l=16$ and $l=8$ overlap perfectly, as shown in the insets of Figs. \ref{figSMNew7}(e)-(g). When $U$ is large, as seen in the inset of Fig. \ref{figSMNew7}(h), the two rescaled profiles separate. We conclude that the STL persists in the presence of weak interaction.

\begin{figure}[hbt]
\includegraphics[width=.95\textwidth]{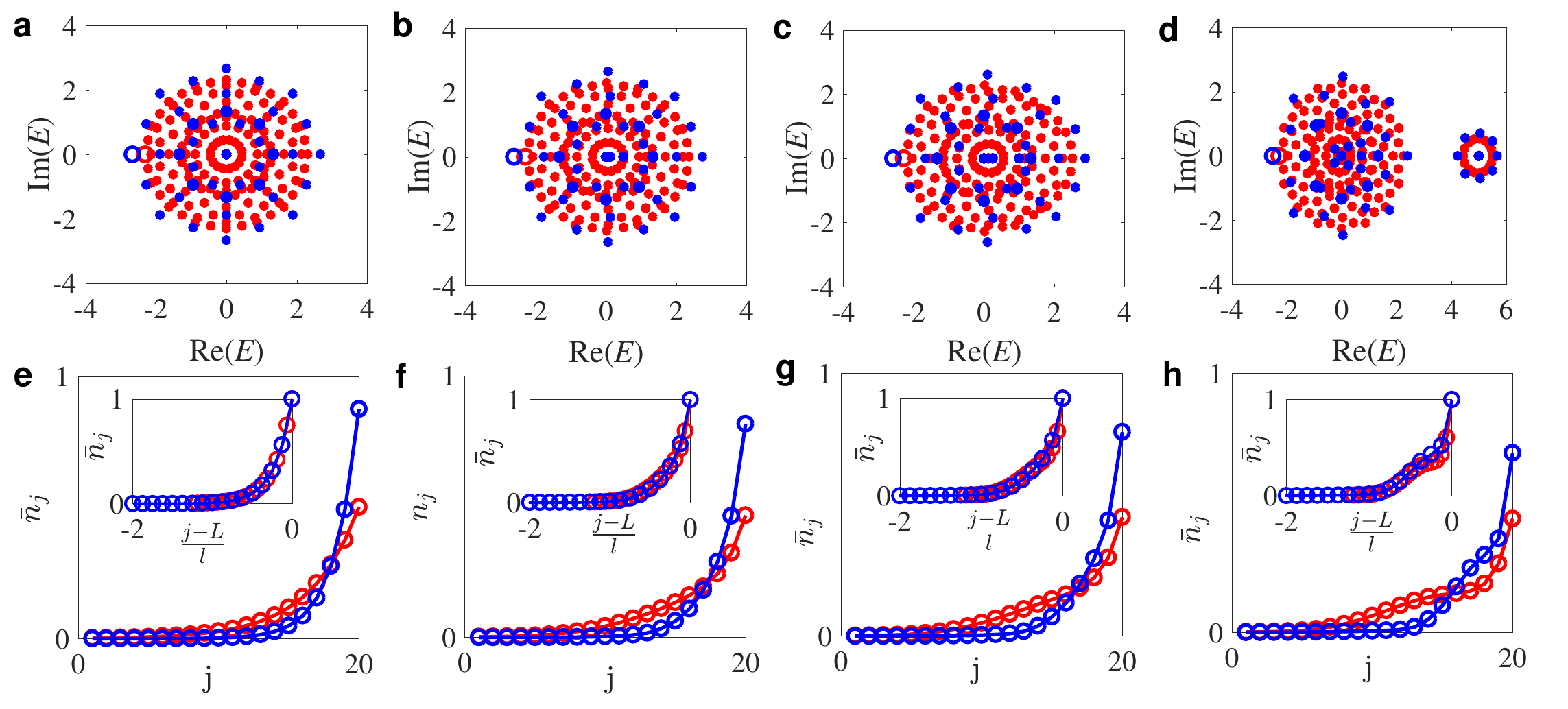}
\caption{(a)-(d) Energy spectra from exact diagonalization for the Bose-Hubbard model with two bosons. The interaction strengths are $U=0$, $U=0.5$, $U=1$, and $U=5$, respectively. The case of $l=16$ or $l=8$ is marked in red or blue. (e)-(h) The spatial profiles of the particle number for the ground state corresponding to the colored circles in (a)-(d). (Insets) The rescaled spatial profiles of particle number for different $l$. Othere parameters: $L=20$, $t=1$, $\delta_t=10$.}
\label{figSMNew7}
\end{figure}

\section*{(VII) Analysis of experimental imperfection or non-uniformity in capacitors and inductors}
In experiments, a unidirectional electrical circuit is employed to observe the STL phenomenon. Inevitably, there are parameter errors (e.g., capacitances and inductances) in the circuit devices during the manufacturing process. We absorb these errors into the admittance matrix $\widetilde{J}$:
\begin{equation}
\begin{split}
&\widetilde{J}_{n,n+1}=-i\omega (C_1+\delta_{C_{1n}}),~~~(n=0,\cdots,N-2)\\
&\widetilde{J}_{N-1,p}=-i\omega (C_2+\delta_{C_{2}}),\\
&\widetilde{J}_{n,n}=-i\omega\delta_{\mu_n},~~~(n=0,\cdots,N-1),\\
\end{split}
\end{equation}
where
\begin{equation}
\begin{split}
&\delta_{\mu_n}=\delta_{C_{1n}}+\delta_{C_{0n}}+\frac{1}{\omega^2L}-\frac{1}{\omega^2(L+\delta_{L_{n}})}, ~~~(n=0,\cdots,N-2)\\
&\delta_{\mu_{(N-1)}}=\delta_{C_{2}}+\delta_{C_{0(N-1)}}+\frac{1}{\omega^2L}-\frac{1}{\omega^2(L+\delta_{L_{N-1}})}.\\
\end{split}
\end{equation}
Here, $\delta_{C_{1n}}$ (or $\delta_{C_{2}}$) represents the deviation from the nominal value of the capacitor that connects the $n$-th (or $(N-1)$-th) node and $(n+1)$-th (or $p$-th) node. $\delta_{C_{0n}}$ (or $\delta_{L_{n}}$) denotes the variation from the nominal value of the capacitor (or inductor) connecting the $n$-th node and the ground. The characteristic polynomial of $\widetilde{J}$ reads
\begin{equation}
\begin{split}\label{Sfj}
f(\tilde{j})=\det\left[\tilde{j}-\widetilde{J}\right]=\prod_{n=0}^{p-1}\left(\tilde{j}+i\omega\delta_{\mu_n}\right)\left[\prod_{n=p}^{N-1}\left(\tilde{j}+i\omega\delta_{\mu_n}\right)-(-i\omega)^l(C_2+\delta_{C_{2}})\prod_{n=p}^{N-2}(C_1+\delta_{C_{1n}})\right].
\end{split}
\end{equation}
Here $\tilde{j}$ denotes the eigenvalue of $\widetilde{J}$, which can be obtained from $f(\tilde{j})=0$. In an ideal scenario without any parameter errors, $\delta_{L_{n}},\delta_{C_{0n}},\delta_{C_{1n}},\delta_{C_{2}}=0$, the eigenvalues can be categorized into two distinct sectors: $p$ eigenvalues merge into $\tilde{j}=0$ corresponding to the $p$-th EP, while the remaining $l$ eigenvalues with $\tilde{j}=-i\omega C_2\sqrt[l]{\frac{C_2}{C_1}}e^{i\frac{2m\pi}{l}}~(m=1,\cdots,l)$ are associated with the scale-tailored localized states. With the device errors taken into account, the eigenvalues related to the $p$-th EP split:
\begin{equation}
\begin{split}
\tilde{j}_n=-i\omega\delta_{\mu_n},
\end{split}
\end{equation}
with $n=0,1,\dots,p-1$. That is, the EP of $\tilde{j}=0$ splits along the imaginary axis of $\tilde{j}$.
\begin{figure}[!h]
\includegraphics[width=0.65\textwidth]{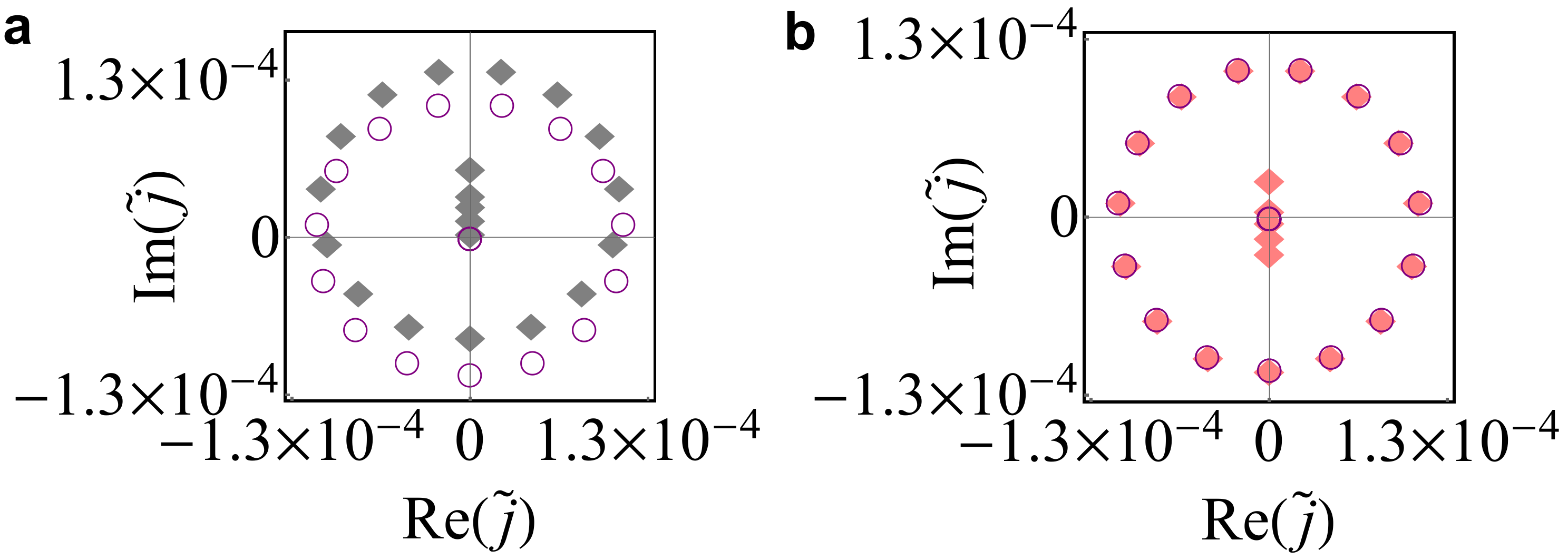}
\caption{Eigenvalues of the admittance matrix $\widetilde{J}$ with $(N,l)=(20,15)$. (a) Theoretical and experimental data are marked by hollow circles and solid diamonds, respectively. (b) Numerical data after considering the device errors of $0.445\%$ for all inductances and $1.5\%$ for all capacitances. Other parameters are: $\omega=2\pi\times100$kHz, $C_0=10$pF, $C_1=220$pF, $C_2=10$pF, $C_3=220$pF, and  $L=220\mathrm{\mu H}$.}
\label{figSM4}
\end{figure}

Figure \ref{figSM4}(a) plots the theoretical and experimental eigenvalues of the admittance matrix $\widetilde{J}$. We then apply a uniform correction of less than $1.6\%$ to all inductances and capacitances in the original data. The numerical results are shown in Fig. \ref{figSM4}(b). The splitting of energy $\tilde{j}$ along the imaginary axis from the $p$-th EP is clearly observed. Additionally, the specific correction data for the capacitances and inductances applied to Fig. 3 of the main text are provided in Table \ref{tableSM1}.
\begin{table}[!h]
\caption{Correction data for device errors associated with inductances and capacitances in Fig. 3 of the main text.}\label{tableSM1}
\begin{center}
\renewcommand{\arraystretch}{1.5}
\begin{tabular}{|c|c|c|c|c|}
\hline
 & $(N,l)=(20,4)$  &  $(N,l)=(20,6)$   &     $(N,l)=(20,9)$  &   $(N,l)=(20,15)$  \\
\hline
$\delta_{L_n}$  &   $0.0052$   &   $0.0052$   &   $0.0058$   &   $0.00445$   \\
\hline
$\delta_{C_{0n}},\delta_{C_{1n}},\delta_{C_{2}}$  &   $0$   &   $-0.01$   &   $0$   &   $-0.015$  \\
\hline
\end{tabular}
\end{center}
\end{table}

\end{document}